\documentclass[useAMS,usenatbib,12pt]{mn2e}
\usepackage[dvipdf]{graphicx,color}
\usepackage[latin1]{inputenc}
\usepackage{graphics}
\usepackage{amsfonts}
\usepackage{amsmath}
\usepackage{multicol}
\usepackage{layout}
\usepackage{amssymb}
\usepackage{epsfig}
\DeclareGraphicsExtensions{.eps}

\title[Galaxy Pair Dynamics]{The Dynamics of Galaxy Pairs in a Cosmological Setting}

\author[J. Moreno et. al.]{Jorge Moreno$^{1,\dagger}$, Asa F. L. Bluck$^{1}$, Sara L. Ellison$^{1}$, David R. Patton$^{2}$, \newauthor Paul Torrey$^{3}$ \&  Benjamin P. Moster$^{4}$ \\
$^{1}$Department of Physics and Astronomy, University of Victoria, Finnerty Road,Victoria, British Columbia, V8P 1A1, Canada\\
$^{2}$Department of Physics and Astronomy, Trent University, 1600 West Bank Drive, Peterborough, Ontario, K9J 7B8, Canada\\
$^{3}$Harvard-Smithsonian Center for Astrophysics, 60 Garden Street, Cambridge, MA, 02138, USA\\
$^{4}$Max Planck Institut f\"{u}r Astrophysik, Karl Schwarzschild Stra\ss e 1, 85748 Garching, Germany \\
$^{\dagger}$CITA National Fellow}

\begin{document}
\date{}
\pagerange{\pageref{firstpage}--
\pageref{lastpage}} \pubyear{2013}
\maketitle
\label{firstpage}


\begin{abstract}

We use the Millennium Simulation, and an abundance-matching framework, to investigate the dynamical behaviour of galaxy pairs embedded in a cosmological context. Our main galaxy-pair sample, selected to have separations $r\leq250\,h^{-1}$ kpc, consists of over 1.3 million pairs at redshift $z=0$, with stellar masses greater than 10$^{9}$ M$_{\odot}$, probing mass ratios down to 1:1000. We use dark matter halo membership and energy to classify our galaxy pairs. In terms of halo membership, central-satellite pairs tend to be in isolation (in relation to external more massive galaxies), are energetically-bound to each other, and are also weakly-bound to a neighbouring massive galaxy. Satellite-satellite pairs, instead, inhabit regions in close proximity to a more massive galaxy, are energetically-unbound, and are often bound to that neighbour. We find that 60\% of our paired galaxies are bound to both their companion and to a third external object. Moreover, only 9\% of our pairs resemble the kind of systems described by idealised binary merger simulations in complete isolation. In sum, we demonstrate the importance of properly connecting galaxy pairs to the rest of the Universe.

\end{abstract}



\section{Introduction}\label{secIntro}

It has been almost fifty years since Arp first published his seminal {\it Atlas of Peculiar Galaxies}, which exemplifies the characteristic effects provoked by interactions and mergers, such as tails, bridges, and other spectacular features \citep{arp66}. In the following decade, the pioneering work of \cite{toomre72} offered a detailed numerical description of the merging process, thus providing a theoretical explanation for Arp's findings \citep[see e.g.,][for an interesting historical review]{duc13}. That same decade also witnessed the emergence of the hierarchical picture of structure formation, which posits that galaxies today are the result of many mergers of smaller units throughout the history of the Universe \citep{ps74,wr78}. Nowadays, mergers play a central role in our understanding of how galaxies form and evolve \citep[see e.g.,][for a recent review]{conselice12}. 

In this context, galaxy pairs are strongly linked to the earliest stages of merging. As such, close pairs have been traditionally used by many to quantify the merger rate of galaxies \citep{carlberg94,patton97,patton00,lefevre00,patton02,bundy04,lin04,depropris05,bell06,kartaltepe07,depropris07,patton08,rawat08,lin08,bluck09,conselice09,deravel09,lopezsanjuan10,bridge10,lin10,lotz11,bluck12,man12,jian12,xu12,lopezsanjuan13}. However, many studies also suggest that galaxy pairs can actually expose the physical effects kindled by the interactions themselves.
 For instance, using the Sloan Digital Sky Survey (SDSS), our group has found that galaxy pairs exhibit enhanced star formation \citep{ellison08,ellison10,patton11,scudder12,ellison13a,patton13}, diluted metallicity \citep{ellison08,scudder12}, and a more prevalent presence of nuclear activity \citep{ellison11,ellison13b}. It is worth mentioning that these effects are captured at projected separations larger than those often considered by merger-rate studies, -- even as far out as $\sim$150 kpc \citep{patton13}.  Similar results have been obtained by other groups \citep{barton00,lambas03,alonso04,nikolic04,kewley06,woods06,alonso07,lin07,barton07,li08,michel08,perez09,freedman10,kewley10,rupke10b,silverman11,wong11,hwang11,ideue12,lambas12,alonso12,sabater13}. 

From a numerical point of view, mergers are also thought to be an integral part of galaxy formation and evolution. For instance, idealised simulations identify galaxy mergers as key agents in (1) shaping the structural properties of galaxies \citep{barnes96,mihos96,naab99,naab03,bournaud04,bournaud05,gonzalez05,boylankolchin05,cox06b,naabtrujillo06,robertson06a,aceves09,hoffman10,bournaud11,bois11,stickley12}; (2) triggering intense episodes of star formation and nuclear activity \citep[][Hayward et al., in prep]{hernquist89,barnes91,mihos94,dimatteo05,hopkins05,hopkins05b,younger09,narayanan10,hayward11,moster11,hopkins12} -- but see \cite{perret13}, whose recent results suggest that merger-induced starbursts are not very frequent at high redshift; and (3) establishing the scaling relations between supermassive black holes and properties of their host galaxies \citep{robertson06b,hopkins07,younger08,johansson09,debuhr11,blecha11,choi13}. More recently, state-of-the-art galaxy merger simulations, capable of approaching parsec-scale physics, have also emerged \citep{teyssier10,hopkins13a,hopkins13b,powell13,renaud13} -- further stressing the importance of this approach in galaxy formation theory. 

In relation to galaxy pairs, a number of simulations pay particular attention to the behaviour of galaxies during the earliest stages of merging. For instance, some works explore how gas turns into stars between first pericentre passage and coalescence \citep{cox06,cox08,dimatteo07,dimatteo08,patton13}, whilst others investigate the nature of chemical evolution as this process unfolds \citep{rupke10,montuori10,perez11,torrey12,scudder12}. These phenomena are attributed to tidal torques, which funnel copious amounts of gas (within the co-rotation radius established by the two galaxies) in to the central regions. Moreover, galaxy encounters can produce gravitational compressive tides, capable of triggering off-nuclear starbursts \citep{renaud08,renaud09}. An additional natural consequence of galaxy-galaxy encounters is the activation of the supermassive black holes at the nuclei of the participating galaxies, and their ultimately coalesce. These interesting phenomena are also captured in galaxy-merger simulations that keep track of the two supermassive black holes \citep{kazantzidis05,callegari09,callegari11,vanwassenhove12,blecha11,blecha13}. 

\begin{figure}
\centering
\includegraphics[width=\hsize]{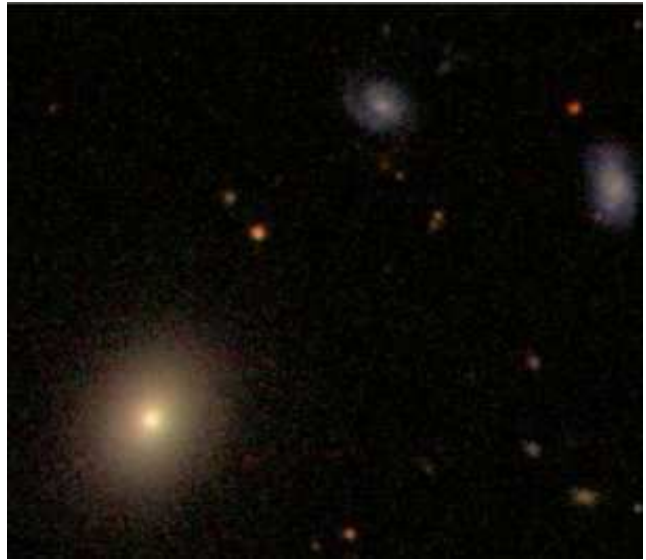}
\caption{A galaxy pair and its surroundings. The upper-right corner shows a pair of spiral galaxies, whilst the lower-left corner shows a massive elliptical galaxy in their vicinity. Image taken from the SDSS SkyServer Object Explorer. Field of view is centred at (Ra,Dec)$=$(220.48,3.395) deg.
}
\label{figsloan}
\end{figure}

In sum, the above results underscore the significance of mergers in the lives of galaxies. Unfortunately, most numerical investigations have a fundamental shortcoming: they rely on {\it the assumption that merging galaxies evolve in isolation}. To illustrate, consider the two spiral galaxies shown at the top right corner of Figure~\ref{figsloan} \citep[this image is drawn from the SDSS close pair catalogue of][]{patton13}. In isolation, one would assume that these galaxies would go through at least one close pericentric passage before merging with one another. However, once the massive elliptical galaxy shown at the bottom left of that Figure is included in the picture, the expected dynamics of the pair are likely to change substantially. Thus, treating galaxy pairs as evolving in complete isolation from the rest of the Universe might not be entirely correct. A central goal of this paper is to use the Millennium Simulation to evaluate the validity of this commonly-adopted assumption.

This paper is organized as follows. In Section~\ref{secMethods} we describe the Millennium Simulation and our abundance matching technique, which we use to construct our main galaxy-pair catalogue. Section~\ref{secResults} shows our results, with particular emphasis on dark matter halo membership and proximity to a third more massive galaxy. In Section~\ref{secDiscussion}, we discuss our sample in the context of binding energy. Our most relevant conclusions are listed in Section~\ref{secConclusions}. 


\section{METHODS}\label{secMethods}

\begin{figure*}
 \centering
    \includegraphics[width=.33\hsize]{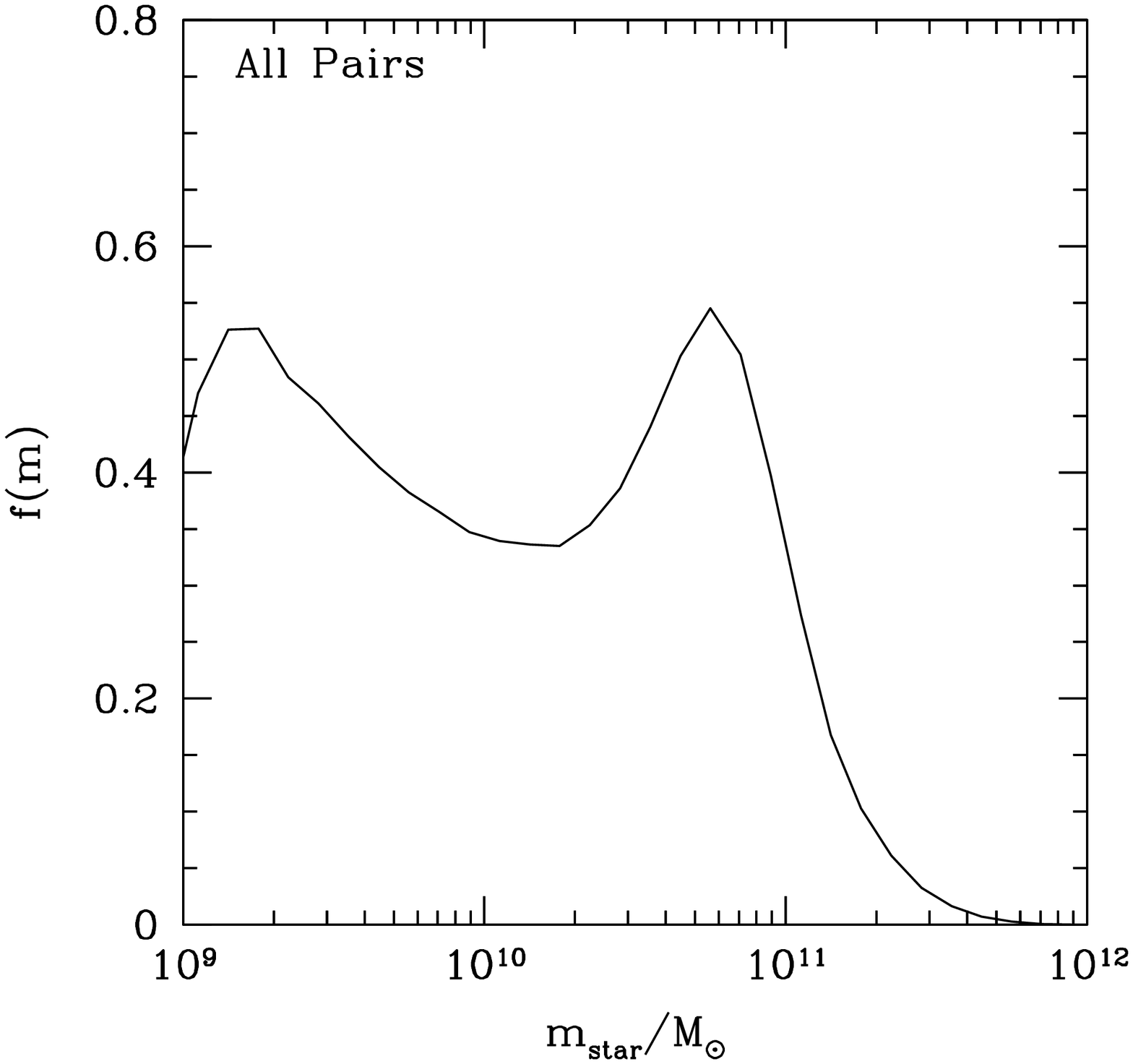}
  \includegraphics[width=.33\hsize]{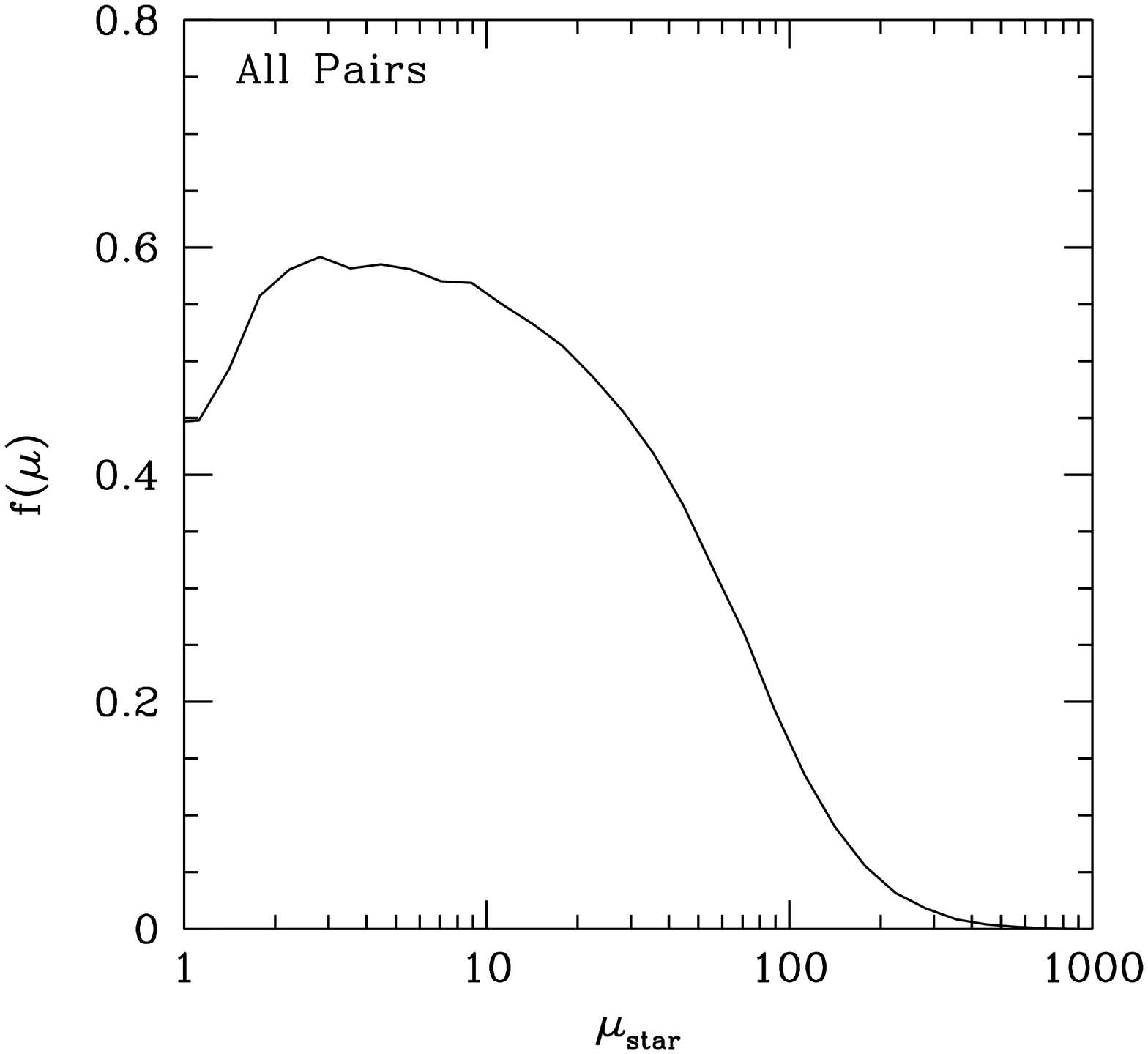}
    \includegraphics[width=.33\hsize]{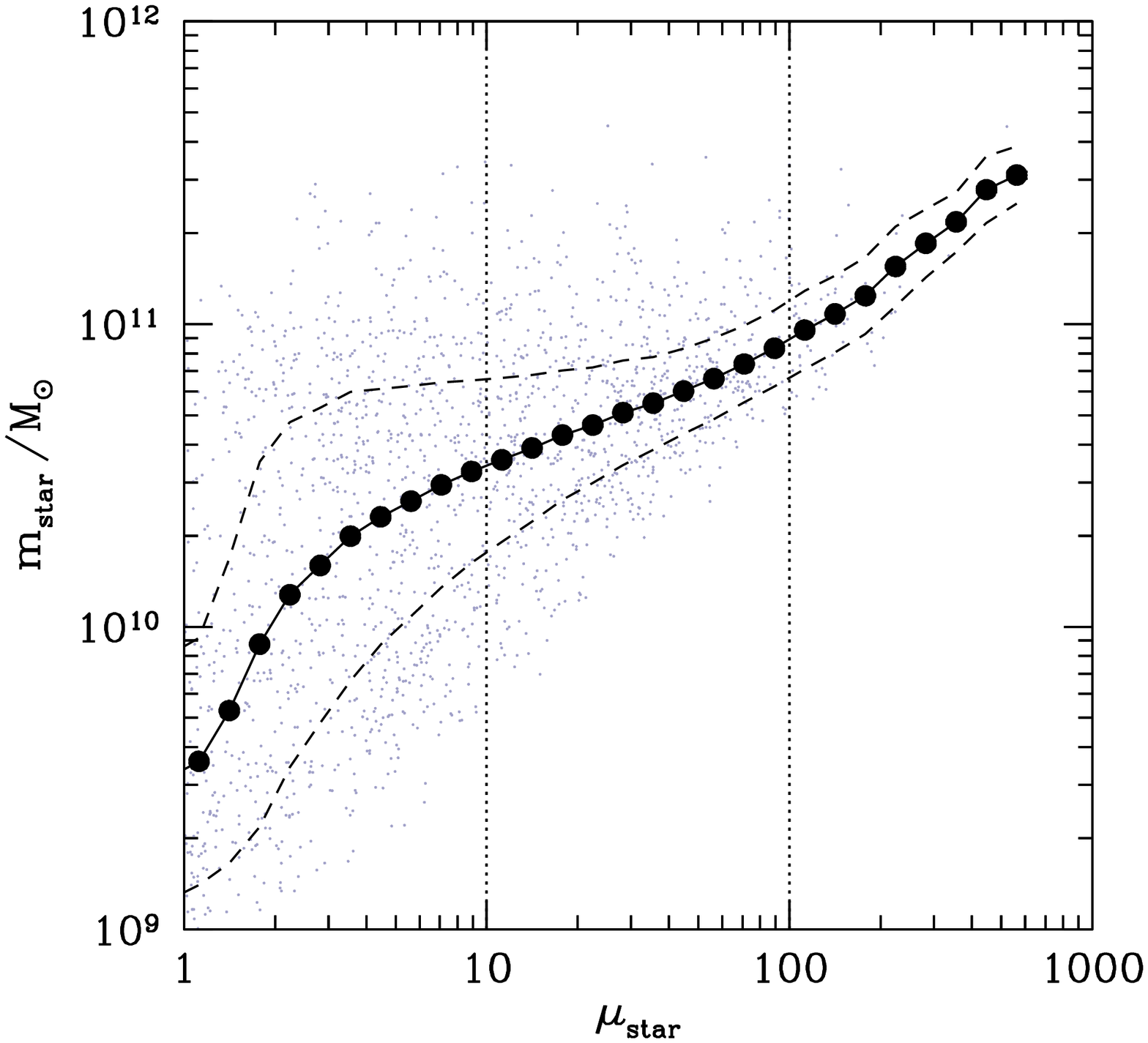}

 \caption{ 
{\it Left}: Histogram of the stellar mass distribution of individual galaxies in our main pair sample. Here, $m_{\rm star}$ refers to the stellar mass of the primary galaxy in the pair (every companion galaxy is counted as a primary somewhere else in the catalogue). {\it Centre:} Distribution of stellar mass ratios. Here, $\mu_{\rm star}$ refers to the ratio of the stellar mass of the more massive galaxy in the pair, to that of its less massive partner (i.e., $\mu_{\rm star} \geq 1$, by definition). {\it Right:} Scatter plot of stellar mass versus mass ratio. The points represent a randomly-selected subsample of our full pair catalogue.  In this figure -- as well as all figures displaying  scatter plots -- filled circles and solid curves represent median values, whilst dashed curves represent 25 and 75 percentiles. The vertical lines demarcate three distinct mass-ratio regimes. The figure shows that, in this sample, pairs with very large mass-ratios tend to require one of the members to be a very massive galaxy.}
 \label{figsample}
\end{figure*}

\subsection{The Millennium Simulation}

In this work we use the Millennium Simulation \citep{springel05}, a dark-matter-only cosmological simulation with $2160^3 \simeq 10^{10}$ particles of mass $m_{\rm p}=8.6 \times 10^{8} h^{-1} M_{\odot} $, in a periodic box of size $L_{\rm box}=500$ $h^{-1}$ Mpc on the side, with Plummer-equivalent force softening scale of 5 $h^{-1}$ kpc, and the following cosmological parameters: $(\Omega_{\rm m},\Omega_{\rm b},\sigma_{8},n_{\rm s},h)=(0.25,0.045,0.9,1,0.73)$. 

In this simulation, gravitationally self-bound structures are identified with the \textsc{subfind} algorithm of \cite{springel01}.  Throughout this work, the term `subhalo' refers to the dark-matter component of either a satellite or a central galaxy. We reserve the term `halo' to the full host dark matter halo, capable of hosting a central galaxy and its satellites. For additional details on the merger history trees, in the context of a Millennium-based catalog of mergers and fly-bys, we refer the reader to \cite{moreno12}.

\subsection{Galaxies: Abundance Matching Technique}

To assign galaxies to subhaloes, we employ the Multi-Epoch Abundance Matching (\textsc{meam}) technique of \cite{moster13}, which uses the latest observed stellar mass functions and the Millennium-based simulated halo mass functions at various redshifts to calibrate the ratio of stellar mass to dynamical (dark-matter subhalo) mass in a galaxy. 

Explicitly, at a fixed redshift $z$, this ratio is parametrized as follows:
\begin{equation}
\frac{m_{\rm star}}{m_{\rm dyn}}\Big|_{\rm z}=2N(z) \Big[\Big(\frac{\bar{m}_{\rm dyn}}{M_{\rm 1}(z)}\Big)^{-\beta(z)}+\Big(\frac{\bar{m}_{\rm dyn}}{M_{\rm 1}(z)}\Big)^{+\gamma(z)}\Big]^{-1},
\label{eqnmratio}
\end{equation}
where the functions $N$, $M_{\rm 1}$, $\beta$ and $\gamma$ are given by equations (11)-(14) of \cite{moster13}, using the parameters quoted in their Table 1. In this expression, 
\begin{equation}
 \bar{m}_{\rm dyn}=\begin{cases}
    m_{\rm dyn}(z) & \text{for central subhaloes},\\
    m_{\rm dyn,\,inf}(z_{\rm inf}) & \text{for satellite subhaloes},
  \end{cases}
\label{eqnmbar}
\end{equation}
where the label `inf' refers to the mass and redshift at which a given satellite was its own central object for the last time (i.e., their values at the {\it infall} time). We note that, in this work, we do not follow `orphan' satellite galaxies, defined as those whose corresponding subhalo is no longer above the resolution limit of the simulation.

In summary, our implementation of the abundance-matching technique assigns stellar masses at infall, which corresponds to the present time for central galaxies (corresponding to redshift $z$), and to earlier times for satellites (corresponding to redshift $z_{\rm inf}>z$). Whilst it is true that the baryonic components of satellite galaxies may be pre-processed between infall and the present \citep[e.g.,][]{abadi99,moore99,balogh00,mccarthy08,bekki09,book10,taranu12,wetzel12,wetzel13}, and possibly even at earlier times \citep{bahe13,mahajan13}; our understanding of these complex physical processes is still under much debate. Therefore, for our present purposes, we omit any extra modelling of the processes affecting satellites during the post-infall period. 

\subsection{The Main Galaxy Pair Sample}\label{secPairs}

We select galaxy pairs in the Millennium Simulation by requiring their comoving three-dimensional separation to be $r \leq 250 h^{-1}$ kpc. This particular threshold is somewhat arbitrary. Though it is chosen to be well-within the one-halo regime prescribed by Halo Occupation Distribution (HOD) models \citep{zheng07} (which is important because ultimately we are interested in capturing those galaxy pairs that are actually interacting). Hereafter, we refer to this catalogue as the {\bf `main' sample}. When appropriate, we also construct sub-catalogues with more stringent proximity thresholds (Section~\ref{seccumd}). 

To avoid numerical issues associated with resolution, we restrict our sample to subhaloes that reach a mass corresponding to at least 100 particles at some point in their history \citep[as done by, e.g.,][]{moreno12}, which is equivalent to 8.6 $\times$ 10$^{10} h^{-1} M_{\odot}$ in dynamical mass (and to approximately 10$^{9} M_{\odot}$ in stellar mass). Also, for this current work, we focus exclusively on redshift $z=0$ (the subject of redshift evolution is deferred to future papers). This results in 1,146,532 individual objects with infall-mass corresponding to at least 100 particles. 

For each mock galaxy satisfying our 100-particle criterion, we construct pairs by finding companions within 250 $h^{-1}$ kpc.  This results in a catalogue consisting of 1,336,289 mock galaxy pairs at $z=0$. We stress that a given simulated galaxy can participate in more than one pair. To illustrate, a group of four galaxies can be potentially split in to six pairs, provided that all their separations satisfy our selection criteria. Other pair catalogues \citep[i.e.,][and previous papers]{patton13} prefer to have at most one pair per galaxy. This is done by taking a galaxy and finding the neighbour with the smallest projected separation (within some line-of-sight velocity difference); whilst ignoring all other potential companions. In this paper, we relax this condition and select all possible pairs, so long as they satisfy our proximity criterion. In this sense, our selection is more in line with studies that make use of correlation-functions \citep{li08,robaina10,robaina12}. 

\begin{figure*}
 \centering
 \includegraphics[width=\columnwidth]{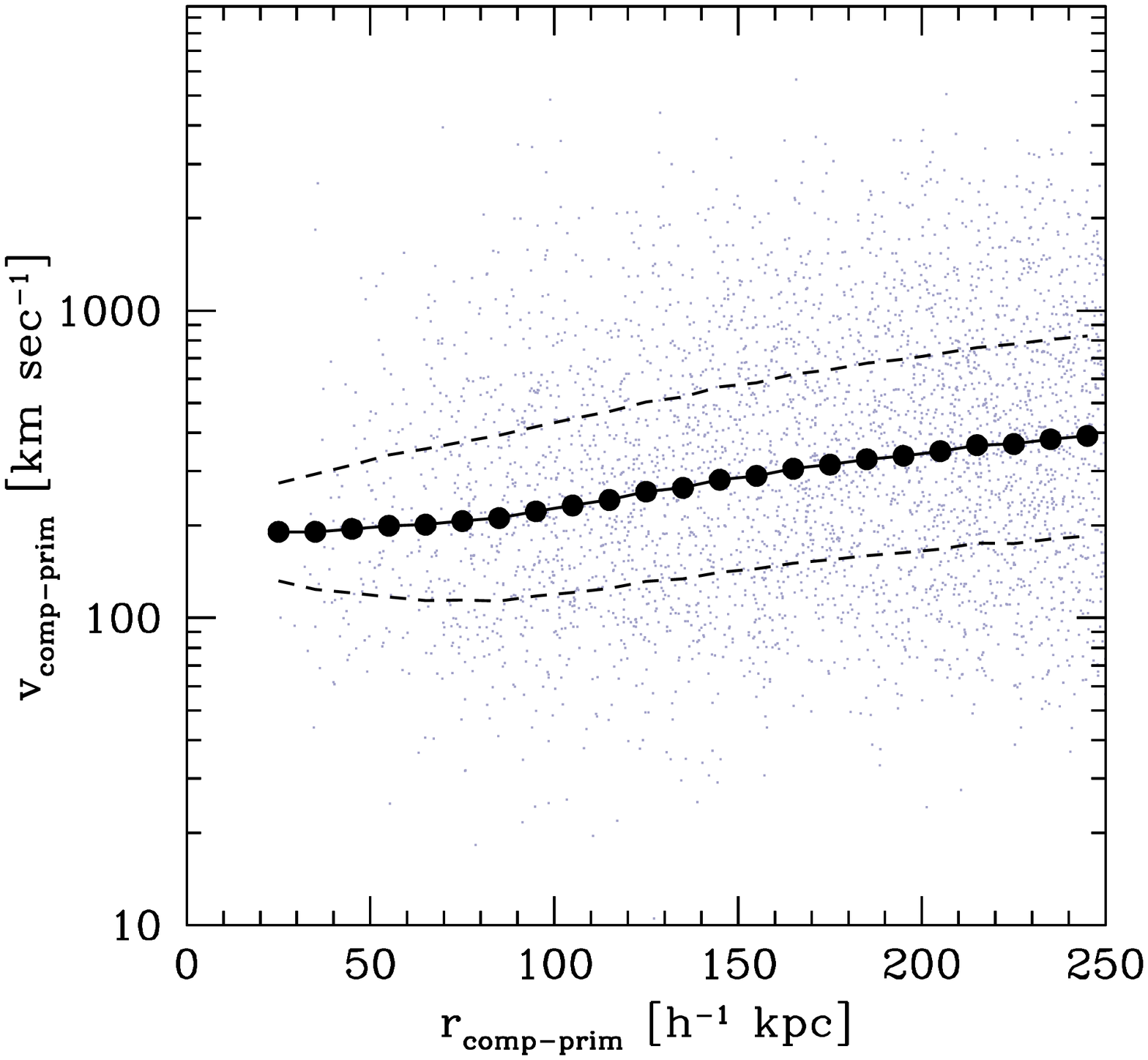}
  \includegraphics[width=\columnwidth]{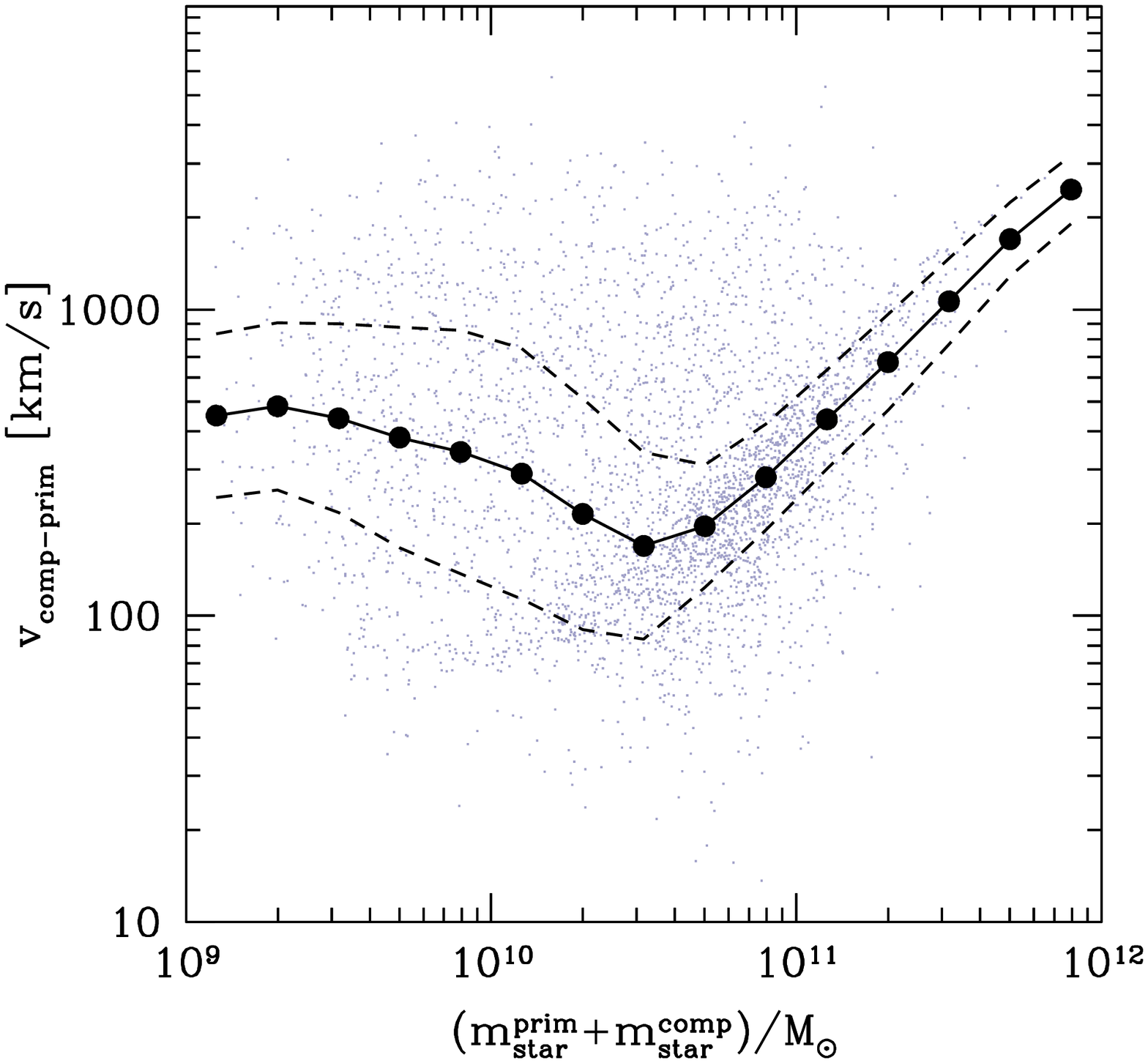}
 \caption{Relative velocity versus separation ({\it left}) and stellar mass ({\it right}). The labels `prim' and `comp' refer to the primary and companion galaxies in the pairs, respectively. Typical relative velocities, and their spread, increase mildly with separation. In terms of mass, the sample appears to be composed of two distinct sub-populations: a diffuse cloud of pairs with low masses and high relative velocities, and a tight sequence of pairs where the relative velocity correlates strongly with mass. Identifying which kind of pairs populate each region of parameter space is one of the goals of this paper.}
 \label{figvdm}
\end{figure*}

Before presenting our results, we briefly discuss the various regimes in stellar mass and stellar-mass ratio accessible to our main sample. The left panel of Figure~\ref{figsample} shows the distribution of stellar masses of individual galaxies in our main pair sample, which ranges roughly from $10^{9} M_{\odot}$ to $\sim10^{12} M_{\odot}$. The explanation of the bimodal nature of this distribution is non-trivial, and we suspect this to be the result of the combination of three effects: (i) the shape of the underlying stellar mass function from which the paired galaxies are drawn; (ii) the behaviour of galaxy clustering as a function of mass; and (iii) the 100-particle limit we impose to avoid low-resolution effects. We do not explore these reasons any further, but keep the shape of this distribution in mind in our forthcoming analysis.

The central panel of Figure~\ref{figsample} shows the distribution of stellar mass ratios, $\mu_{\rm star}$, computed by taking the ratio of the stellar mass of the dominant galaxy in the pair to that of its smaller companion. Thus, by definition, $\mu_{\rm star} \geq 1$. The distribution dips at low values of $\mu_{\rm star}$ because pairs of galaxies with nearly identical masses are less common than those with widely discrepant masses. At high values of $\mu_{\rm star}$, the distribution tails off as well. This is explained as follows: the finite mass resolution of the simulation implies that in order to reach very high mass ratios, one also needs the larger galaxy in the pair to be very massive. However, the finite volume of the simulation restricts the number of very massive galaxies available, thereby limiting the number of pairs with extreme mass ratios as a result. It is interesting to note that observations also suffer from increasing incompleteness at large $\mu_{\rm star}$. However, there the main cause is the fact that observed samples are usually magnitude limited.

Lastly, the right-hand panel of Figure~\ref{figsample} describes the interplay between stellar mass and mass ratio, which shows a scatter plot of these two quantities. For illustration purposes, the points depict a randomly-selected subset of our sample. The black solid symbols and the solid curves show the median, and the dashed curves show the 75 and 25 percentiles, respectively. (A number of scatter plots in this paper will follow this format.) As stated above, the stellar masses and mass ratios are correlated, especially at large values of both parameters. This occurs because the simulation is mass-limited (i.e., the lack of points on the lower-right corner) and volume limited (i.e., the lack of points above a horizontal line at high $\mu_{\rm star}$). In other words, `major' pairs (i.e., those with low values of $\mu_{\rm star}$) exist for a broad range of stellar masses. On the other hand, to probe galaxy pairs with a wide range of mass ratios, very large values of stellar mass are required. Lastly, we note that a hard diagonal cut across the $M_{\rm star}-\mu_{\rm star}$ plane (right panel of Figure~\ref{figsample}) is not evident because we make our selection cut at $m_{\rm dyn,\,inf}=100 \times m_{\rm p}$, which does not translate directly into a single value of $m_{\rm star}$ at $z=0$. This is because satellites at that epoch have a range of $z_{\rm inf}$ values (equation~\ref{eqnmbar}).


\section{Results}\label{secResults}

\subsection{Global Dynamical Trends}\label{secCosmo}

Let us now explore the nature of our simulated main pair catalogue in the context of dynamics. The left panel of Figure~\ref{figvdm} shows the velocity of a galaxy in a pair (hereafter labeled `prim' for {\bf primary}) relative to its {\bf companion} (labeled `comp'). In our terminology, there is no fundamental difference between a primary and a companion galaxy. In fact, the catalogue is built so that every galaxy with a neighbour within 250 $h^{-1}$ kpc is considered a primary. Thus, by construction, every companion in that neighbourhood is also regarded as a primary somewhere else in the list (with the original primary now taking the place of the companion at that later slot in the list).  

To illustrate, consider the Milky Way (MW) and the Large Magellanic Clouds (LMC). In this case, the MW would the primary in the MW-LMC pair, whilst the LMC would the companion. The roles are reversed for the LMC-MW pair, where the LMC is now considered the primary and the MW becomes the companion. Notice that, with our convention, every galaxy pair appears twice in our catalogue -- which is of no statistical consequence to our analysis.

Also, we reiterate that a given galaxy can be a member of more than one pair. Namely, we assign one pair for every companion within our proximity threshold. To illustrate, image pairs involving the MW, the LMC and the Small Magellanic Cloud (SMC). In our scheme, the MW would belong to {\it both} the MW-LMC pair and the MW-SMC pair -- whilst other authors would be forced to choose between the LMC, the SMC, or some other galaxy, as the `unique' companion of the Milky Way.

One must keep in mind that this sample includes a wide array of stellar masses, mass ratios, environments and stages of interaction -- making the interpretation of the upcoming plots somewhat non-trivial. In particular, notice that in Figure~\ref{figvdm} (left panel), the median of velocity (solid symbols and curve) increases with separation, and so does the spread, which is demarcated by the 25 and 75 percentiles (dashed curves).  Also note that many of these three-dimensional pairs have relative velocities larger than 1,000 km sec$^{-1}$, substantially larger than those often assumed by observational works \citep[i.e., of the order of 300 km sec$^{-1}$, using line-of-sight velocities,][]{patton13}. Nevertheless, in the following sections (i.e., the right panel of Figure~\ref{figvdf}, the left panel of Figure~\ref{figqphys}, and associated text) we argue that the velocity cuts imposed by observers are not entirely misguided.

The right panel of Figure~\ref{figvdm} shows the relative velocity again, but now against the total stellar mass of the pair. In the high stellar-mass regime (i.e., $\gtrsim 3\times10^{10} M_{\odot}$), the points delineate a strong correlation between relative velocity and total stellar mass. In contrast, the low-mass regime materializes as a diffuse cloud containing galaxies moving at rather large velocities relative to their companions. It is natural to expect that, on average, the magnitude of the relative velocity should scale with the mass of the system. Namely, the more massive the galaxies in the pair, the deeper the gravitational potential which, on average, is capable of driving larger relative velocities. However, we see that our sample contains a large portion of pairs that do not follow that simple rule. 

One possibility is that these low-mass/high-velocity pairs are not the main drivers of their own dynamics. Instead, it is likely that these consist of pairs of satellite galaxies orbiting a massive dark matter halo. In other words, in a large fraction of cases, it might not suffice to consider pairs of galaxies as if they were evolving in complete isolation. Instead, it is imperative to explore their vicinity and check if these pairs belong to a more massive system responsible for driving their dynamics. Indeed, as Figure~\ref{figsloan} seems to suggest, this might be the case -- which we demonstrate next.

\subsection{Halo Membership: Galaxy Pair Flavours}\label{secFlavs}

\subsubsection{Galaxy Pair Flavours: Definitions}\label{secflavdefs}

In this section we investigate the role of the host dark matter halo(es) in driving the dynamics of our simulated galaxy pairs. 
Several models of galaxy formation \citep[e.g.,][and reference therein]{sales07,delucia12} suggest that central and satellite galaxies in a given dark matter halo experience different evolutionary paths. In this context, a {\it central galaxy} refers to the most massive galaxy at the centre of the potential. In galaxy clusters, this would correspond to the brightest cluster galaxy (BCG). Some authors argue that the central galaxy in a group need not be the brightest galaxy \citep{skibba11}. Here we follow \cite{delucia07}, who use the term `central' galaxy to refer to the dominant galaxy in the history of a dark matter halo. {\it Satellite galaxies}, on the other hand, were accreted later on and, if identified at $z=0$, they are still orbiting the potential well of their host dark halo.\footnote{Explicitly (i.e., for Millennium Simulation users), {\it central} subhaloes are those with \textsf{haloID}$=$\textsf{firstHaloInFOFgroupId}, whilst {\it satellite} subhaloes are those with \textsf{haloID}$\neq$\textsf{firstHaloInFOFgroupId}.} 

Models show that these galaxies tend to be more susceptible to environmental effects, such as ram-pressure stripping, etc. \citep[e.g.,][and references therein]{wetzel13}. This assertion has observational support as well \cite[see e.g.,][Bluck et al., in prep]{peng12,kovac13}.
 
That being said, the language of central and satellite galaxies allows us to classify pairs into {\it five} distinct {\bf flavours}:

\begin{itemize}

\item[(a)] {\bf Central-Satellite (1 Halo)}: As the name indicates, these are pairs connecting the central galaxy in a dark matter halo to one of its satellites. A pair involving the MW and the LMC are a prototype example of this. Similarly, a pair involving a BCG and a nearby satellite also counts towards this population \citep[e.g.,][]{edwards12}.\\

\item[(b)] {\bf Satellite-Satellite (1 Halo)}: A pair of two satellite galaxies in the same dark matter halo. The LMC and the SMC make a good example for this type of pair. Satellite-satellite pairs have also been identified in galaxy clusters \citep{vandokkum99,tran05}. \\

\item[(c)] {\bf Central-Central (2 Haloes)}: A pair connecting two central galaxies in distinct dark matter haloes. Two galaxies analogous to the MW-M31 pair would qualify. (However, the distance between the MW and M31 is too large for this particular pair to pass our selection criteria.)\\

\item[(d)] {\bf Central-Satellite (2 Haloes)}: A pair connecting a central galaxy to a satellite of a {\it different} halo. To illustrate, imagine a small group on the verge to being accreted by a nearby massive cluster. In some cases, it is plausible for the central galaxy of that group to be physically close to one of the satellites sitting on the outskirts of the cluster.\\

\item[(e)] {\bf Satellite-Satellite (2 Haloes)}: A pair connecting two satellite galaxies in {\it distinct} haloes. For example, consider two rich clusters right on the brink of merging. At this stage, two satellites belonging to each of the two clusters could potentially create a galaxy-pair configuration.\\
\end{itemize}
 
\begin{figure}
 \centering
 \includegraphics[width=\hsize]{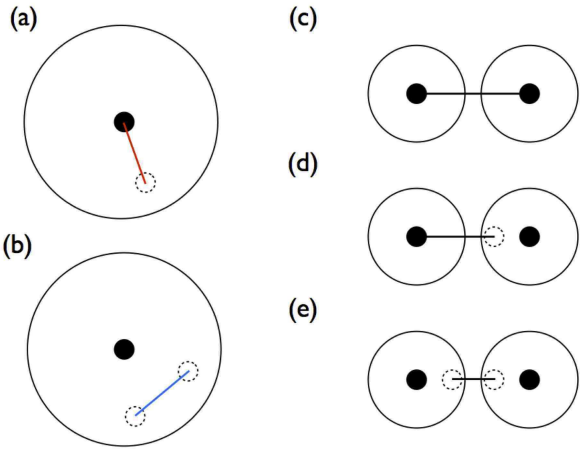}
 \caption{Galaxy pair flavours: definitions. In the context of central and satellite galaxies, pairs are classified as follows: (a) central-satellite (one halo), (b) satellite-satellite (one halo), (c) central-central (two haloes), (d) central-satellite (two haloes) and (e) satellite-satellite (two haloes). For flavours (a) and (b), we connect the pairs with coloured lines: red and blue, respectively. This particular choice of colours is kept throughout the paper. See Table~\ref{tableflavs} for the relative contributions from each flavour.}
 \label{figflavs}
\end{figure}

\begin{center}
\begin{table}
\begin{tabular}{|| l |l l || c || l ||}
\hline
 & Flavour & Percentage & Colour \\
\hline \hline 
--- & All Pairs & 100 \% & black \\
(a) & Central-Satellite (1 Halo)  & 52.7 \% & red \\
(b) & Satellite-Satellite (1 Halo)  & 46.1 \% & blue\\
(c) & Central-Central (2 Haloes)  & 1.1 \% & -- \\
(d) & Central-Satellite (2 Haloes)  & 0.15 \% & -- \\
(e) & Satellite-Satellite (2 Haloes)  & 0.003 \% & -- \\
\hline
\end{tabular}
\caption{Galaxy pair flavours: relative contributions. Each row refers to a galaxy pair flavour, as depicted in Figure~\ref{figflavs}. Flavour name, percentage contribution to $z=0$ pair population, and colour used in figures, are given in the columns. Due to their small contribution at $z=0$, we limit the discussion of flavours (c)-(e), and focus exclusively on flavours (a) and (b) in this paper.}
\label{tableflavs}
\end{table}
\end{center}

For reference, Figure~\ref{figflavs} elucidates the meaning of the various {\it flavours} in a schematic fashion: large circles represent host dark matter haloes, filled circles represent central galaxies, dotted circles refer to satellite galaxies, and the solid lines connect pairs of galaxies. Table~\ref{tableflavs} reports the relative contribution of each flavour. We remind the reader that, in our catalogue, any given galaxy can participate in more than one pair. For example, the LMC can be considered to be a member of both the MW-LMC pair (flavour a) and the LMC-SMC pair (flavour b).

We now take a moment to discuss the relative numbers in each flavour, which depend strongly on our choice of pair selection criteria, the regimes the simulation can probe, and the fact that we are focusing our attention to $z=0$ in this paper. 
First, notice that in our sample, the great majority of pairs (almost 99\% of them) inhabit the same dark matter halo (flavours a and b). Had we chosen a sufficiently larger critical separation in our selection criteria (much larger than the typical dark matter halo diameter at this redshift), our sample would include more central-central pairs (flavour c). Similarly, at high $z$, where the typical halo diameters are smaller (there are fewer cluster-sized haloes), we expect to find more central-central pairs. 

Also, a simulation with higher resolution (and thus with more low-mass haloes with smaller diameters) would include more central-central pairs \citep[i.e., Millennium-II,][]{boylankolchin09}. However, such simulation would also include more substructures, and thus more central-satellite and satellite-pairs. At this stage, it is unclear how the relative contributions to the various adopted flavours would change with a different simulation -- here we just wish to stress that these relative numbers depend on the resolution of the simulation.

\begin{figure*}
 \centering
  \includegraphics[width=\columnwidth]{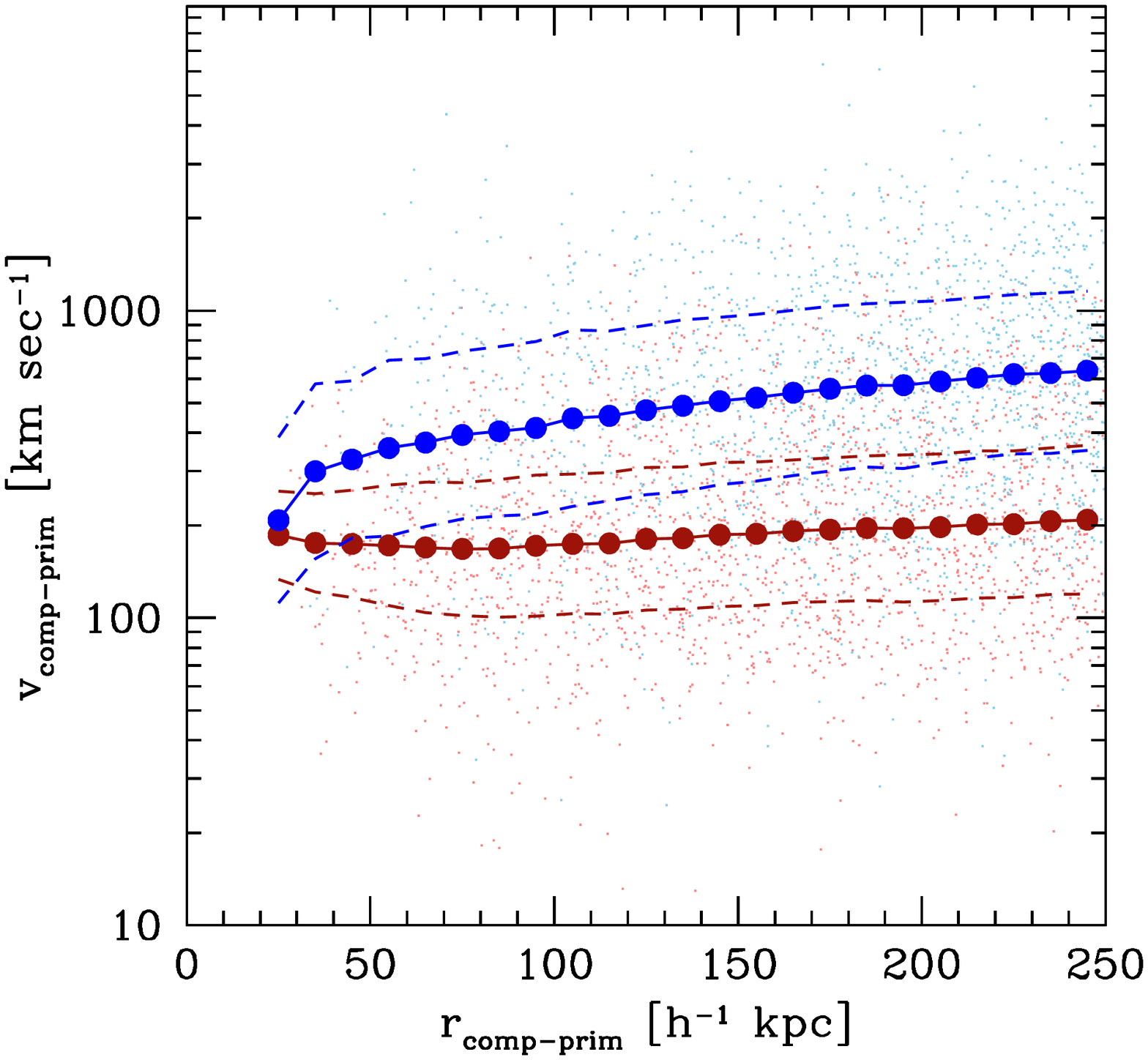}
  \includegraphics[width=\columnwidth]{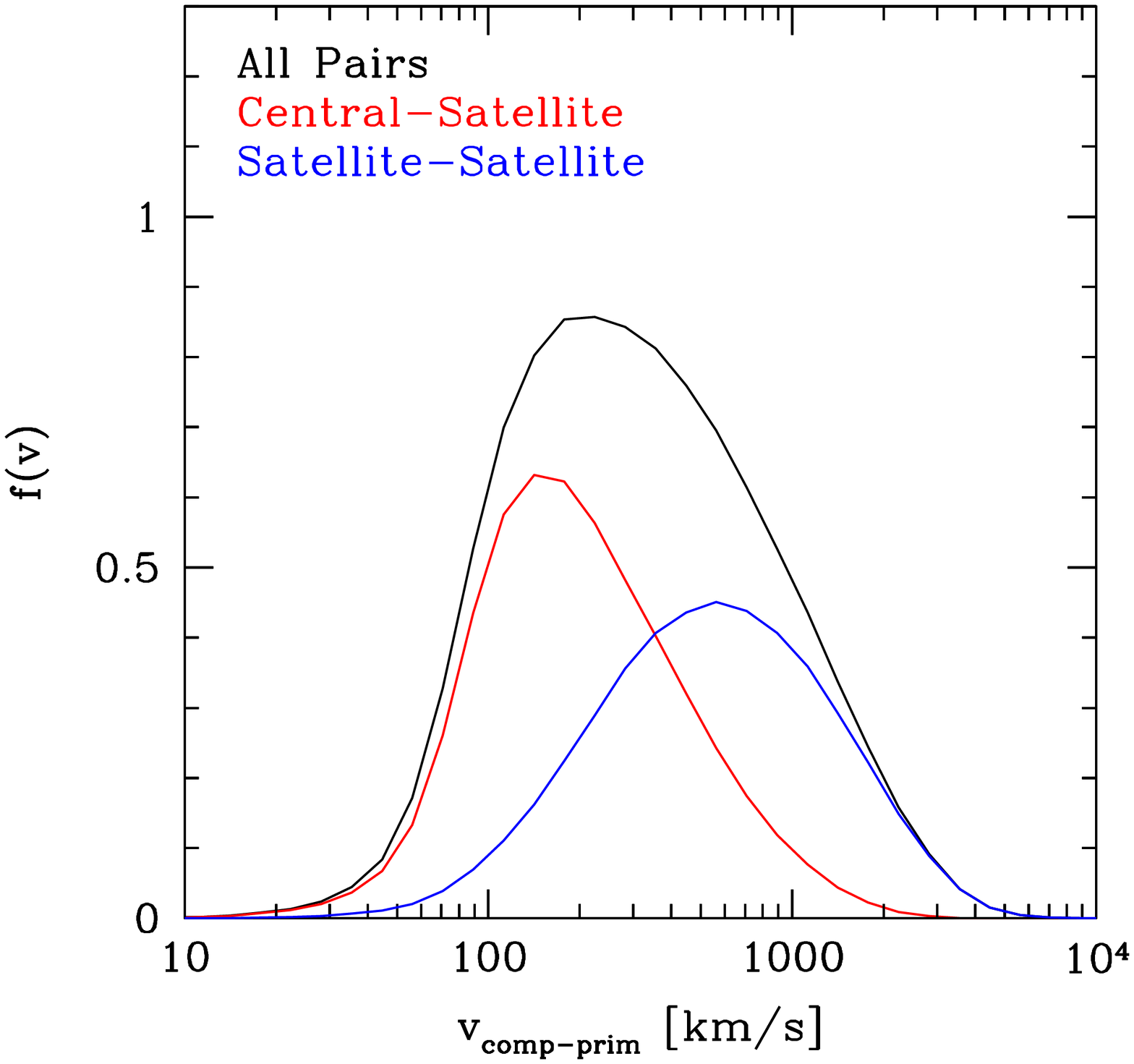}

 \caption{{\it Left}: Relative velocity versus separation for central-satellite pairs (red points) and satellite-satellite pairs (blue points). The other three flavours always have separations larger than 150 $h^{-1}$ kpc and, due to their low numbers, are omitted for the rest of the paper (see Figure~\ref{figflavs} and Table~\ref{tableflavs} for definitions and relative contributions).  The dependence of relative velocity on separation is negligible for central-satellite pairs, and weak for satellite-satellite pairs. {\it Right}: Distribution of relative velocities. Here and throughout the rest of the paper, black curves correspond to the entire galaxy pair sample, whilst red and blue curves represent the contributions from central-satellite and satellite-satellite pairs, respectively.  Both panels show that, typically, satellite-satellite pairs experience higher relative velocities than their central-satellite counterparts (i.e., peaking at $\sim$600 km sec$^{-1}$ versus $\sim$200 km sec$^{-1}$).
  }
 \label{figvdf}
\end{figure*}

In principle, the evolution of the relative flavour contributions with redshift can be explained in terms of the orbits the galaxies experience, and their merger history.  For instance, imagine a central-satellite pair in a given dark matter halo (flavour a). If we track the satellite galaxy back in time, we find that at the infall epoch, the progenitor of this pair was actually a central-central pair (flavour c). Similarly, the members of a satellite-satellite pair (flavour b) can be tracked in time to epochs where this pair was a central-satellite pair (flavour a) in the outskirts of a massive halo -- and even further back to when this pair was a central-central pair (flavour c) at high $z$. Of course, it is possible that these two galaxies were just paired up recently, and were in fact isolated galaxies in the past. The analysis of all these interesting configurations is the subject of an upcoming paper (Moreno et al., in prep).

Lastly, we comment on the infrequency of galaxy pairs involving two haloes (in our $z=0$ sample), which increases as one moves from flavour (c) all the way down to flavour (e).
This can be understood as follows. A central-central pair (flavour c) typically requires low-mass haloes, so that the galaxy-galaxy separation is larger than the mean of the diameters of the two host haloes. Often, due to the resolution of the simulation, many of these low-mass haloes contain only a single central galaxy. However, the situation is different for two-halo central-satellite pairs (flavour d). In this case, the halo containing the satellite has to be massive enough (and therefore large enough) to contain satellites in the first place (above the resolution of the simulation). The same is true for two-halo satellite-satellite pairs (flavour e), except that we now require both haloes to be massive enough (and large enough) to host satellites. Given that, in our hierarchical picture, there are far more low-mass haloes (i.e., with a single galaxy above the resolution of the simulation) than those with satellites -- this renders flavour (d) and particularly flavour (e) far less infrequent than flavour (c). 

For the remainder of this work, we will ignore galaxy pairs involving two distinct haloes (flavours c, d and e). However, we keep this terminology for future papers -- where we are investigating the dynamics of pairs selected at high $z$, the progenitors of pairs selected at low $z$, and explore modifications to our selection criteria (Moreno et al., in prep).
 
\subsubsection{Galaxy Pair Flavours: Dynamics}\label{secflavdyn}

We now focus our attention to the dynamical behaviour of galaxy pairs using the notion of flavours. Figure~\ref{figvdf} shows relative velocities versus separations in our pair catalogue. This scatter plot is analogous to the left panel of Figure~\ref{figvdm}, except that the sample is now split into its two dominant flavours: red points refer to central-satellite pairs, whilst blue points refer to satellite-satellite pairs. (These colour choices are kept throughout a number of the forthcoming figures.) 

For central-satellite pairs, relative velocities are largely independent of separation. Satellite-satellite pairs, on the other hand, tend to exhibit a slight increase in relative velocity with larger separations.  On average, at fixed separation, satellite-satellite pairs tend to have larger relative velocities than their central-satellite counterparts. This trend is confirmed by the histograms presented on the right-hand panel of Figure~\ref{figvdf}. This figure clearly shows that the distribution of relative velocities peaks at a lower value for central-satellite pairs ($\sim$200 km sec$^{-1}$) than for satellite-satellite pairs ($\sim$600 km sec$^{-1}$). In particular, notice that the strict velocity cut at 300 km sec$^{-1}$ commonly imposed by observers \citep[e.g.,][line-of-sight component]{patton13} lies right in between the two peaks. In other words, such a velocity cut has the potential to discard a large fraction of satellite-satellite pairs, whilst keeping a large fraction of central-satellite pairs. This is relevant because later in the paper (Section~\ref{secEnergy}) we demonstrate that the majority of central-satellite pairs are in fact bound (see also right panel of Figure~\ref{figqphys} and discussion therein).

\begin{figure*}
 \centering
  \includegraphics[width=\columnwidth]{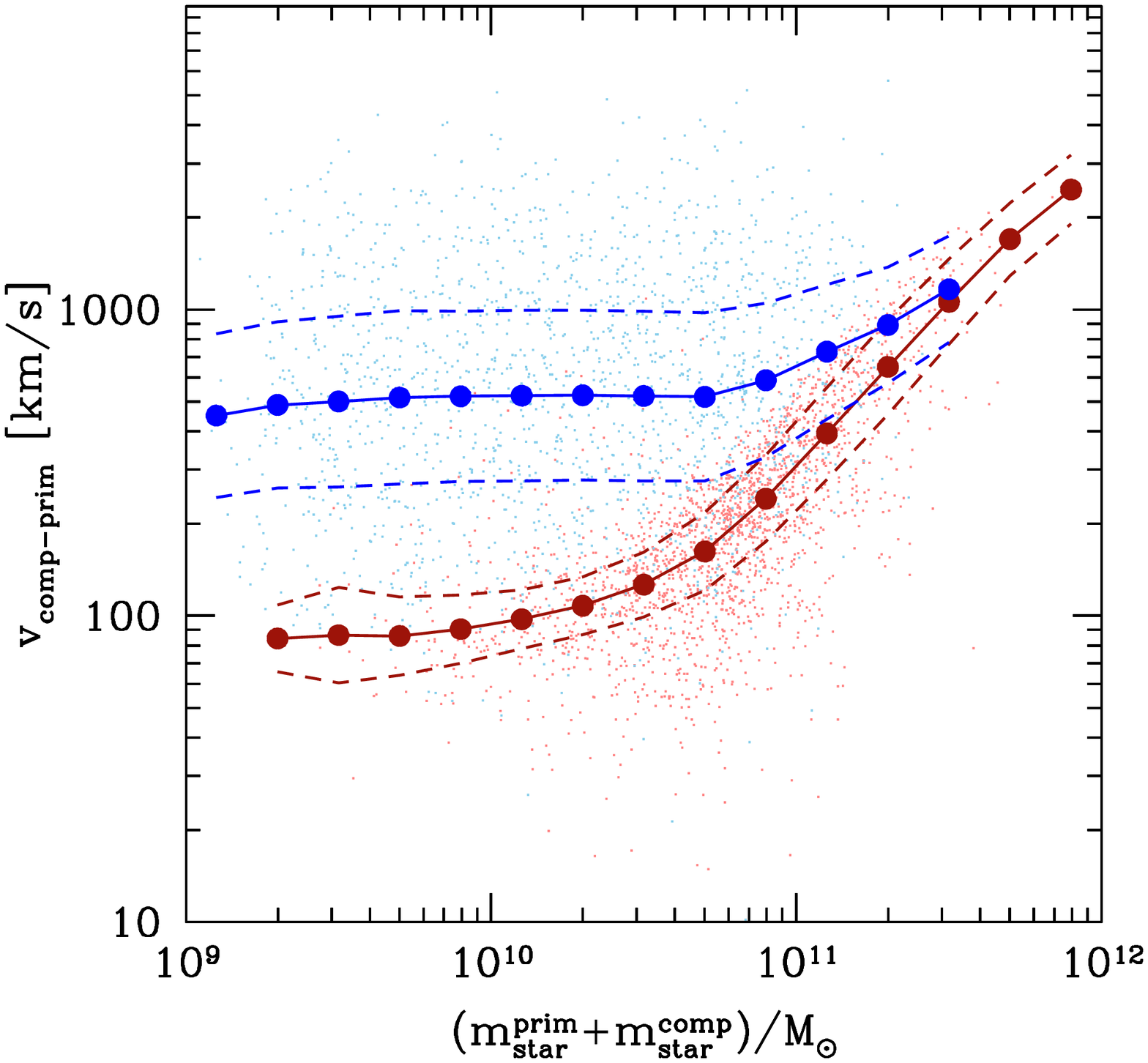}
  \includegraphics[width=\columnwidth]{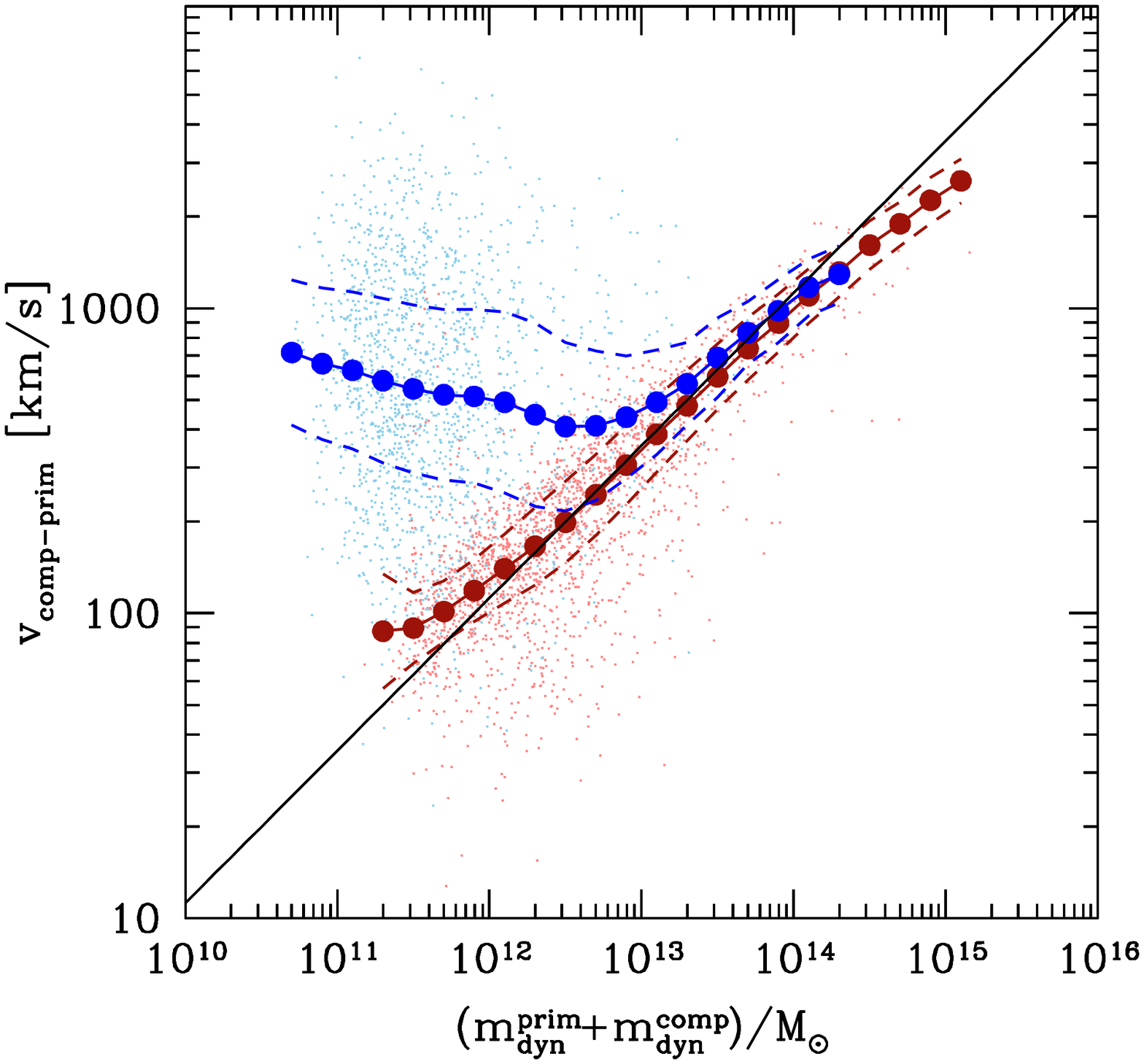}

 \caption{Relative velocity versus total stellar mass ({\it left}) or total dynamical (dark matter) mass ({\it right}), for central-satellite pairs (red points) and satellite-satellite pairs (blue points).  These two panels demonstrate that galaxy pairs can be naturally divided into two sub-populations: (i) central-satellite pairs, whose relative velocities scale strongly with total stellar mass; and (ii) satellite-satellite pairs, consisting of primarily low-mass systems whose relative velocities are high and do not scale with mass, except for the most massive galaxy pairs. In particular, the mass-velocity trend followed by central-satellite pairs is more evident in the right-hand panel because it is gravity (which is connected to dynamical mass) the factor that drives this relation. We place the diagonal solid line $
v=\sqrt{2G(m^{\rm prim}_{\rm dyn}+m^{\rm comp}_{\rm dyn})/(250 \,h^{-1}\,{\rm kpc})}$ to guide the eye. 
 }
 \label{figvmf}
\end{figure*}

The left panel of Figure~\ref{figvmf} shows a scatter plot of relative velocity versus total stellar mass. This figure is analogous to the right-hand panel of Figure~\ref{figvdm}, but now splits the sample into central-satellite pairs (red points) and satellite-satellite pairs (blue points). As we anticipated in Section~\ref{secflavdefs}, {\it central-satellite pairs exhibit a strong correlation between relative velocity and total stellar mass, whilst satellite-satellite pairs generally do not}.  In other words, central-satellite pairs drive their own dynamics, whilst the dynamics of satellite-satellite pairs is governed by their host dark matter halo.

Notice that for central-satellite pairs, the velocity-mass correlation seems to flatten out as we approach the low-mass regime (i.e., below $\sim 2 \times 10^{10} M_{\odot}$). The reason behind this is that, at low masses, a relatively small bin in stellar mass encompasses a rather large bin in dynamical mass. In Figure~\ref{figvmf}, we show the relative velocities versus total dynamical (subhalo) masses. We include a diagonal line at $
v=\sqrt{2G(m^{\rm prim}_{\rm dyn}+m^{\rm comp}_{\rm dyn})/(250 \,h^{-1}\,{\rm kpc})}$ to guide the eye (here, $G$ denotes Newton's gravitational constant). In terms of dynamical masses, the velocity-mass correlation in central-satellite pairs not only becomes more evident, but so does the bifurcation of the sample by flavour. Moreover, the `squeezing' effect due to shifting our scatter plot from $m_{\rm dyn}$ to $m_{\rm star}$ is no longer present in that panel.

\begin{figure*}
 \centering
    \includegraphics[width=\columnwidth]{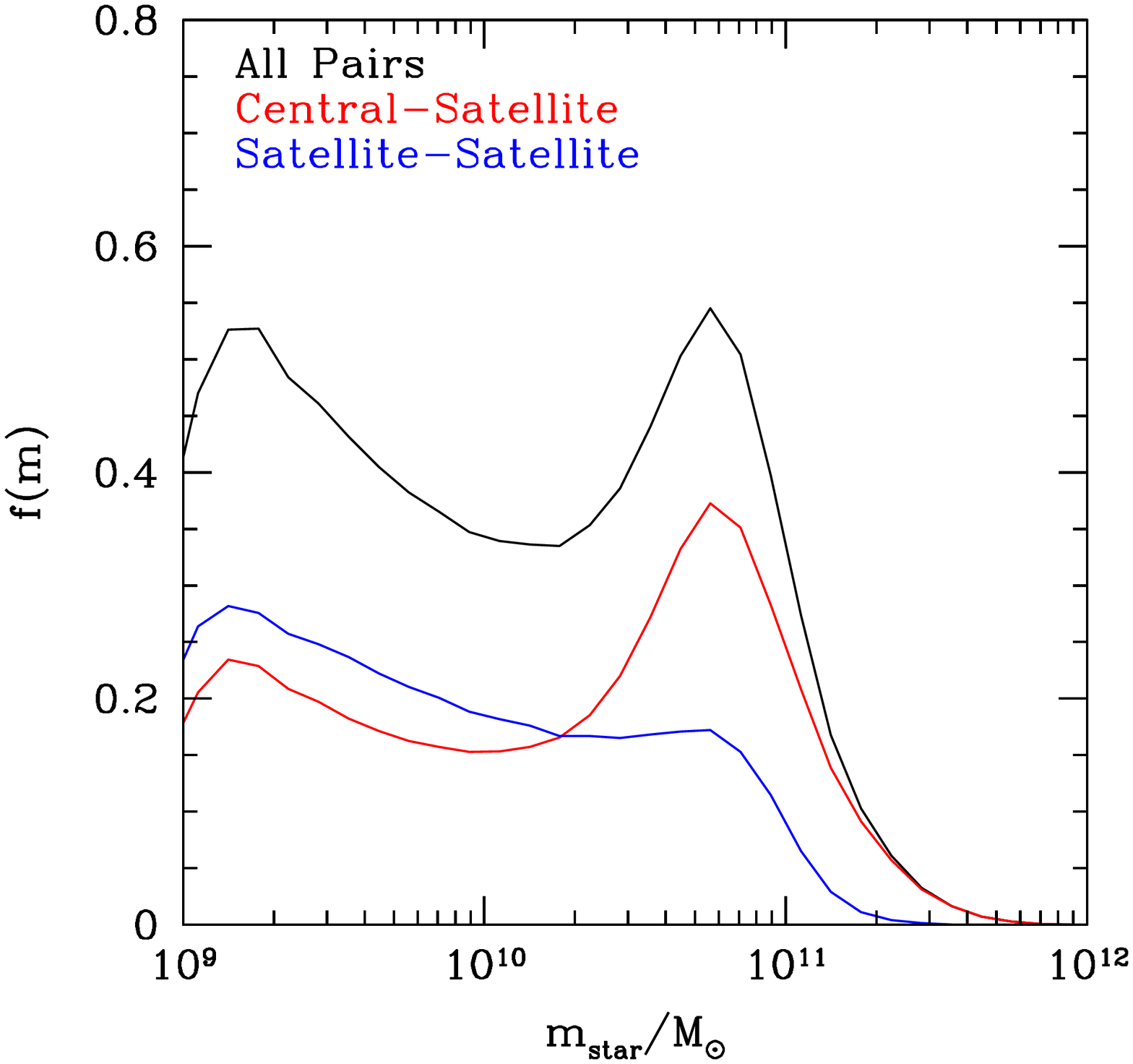}
  \includegraphics[width=\columnwidth]{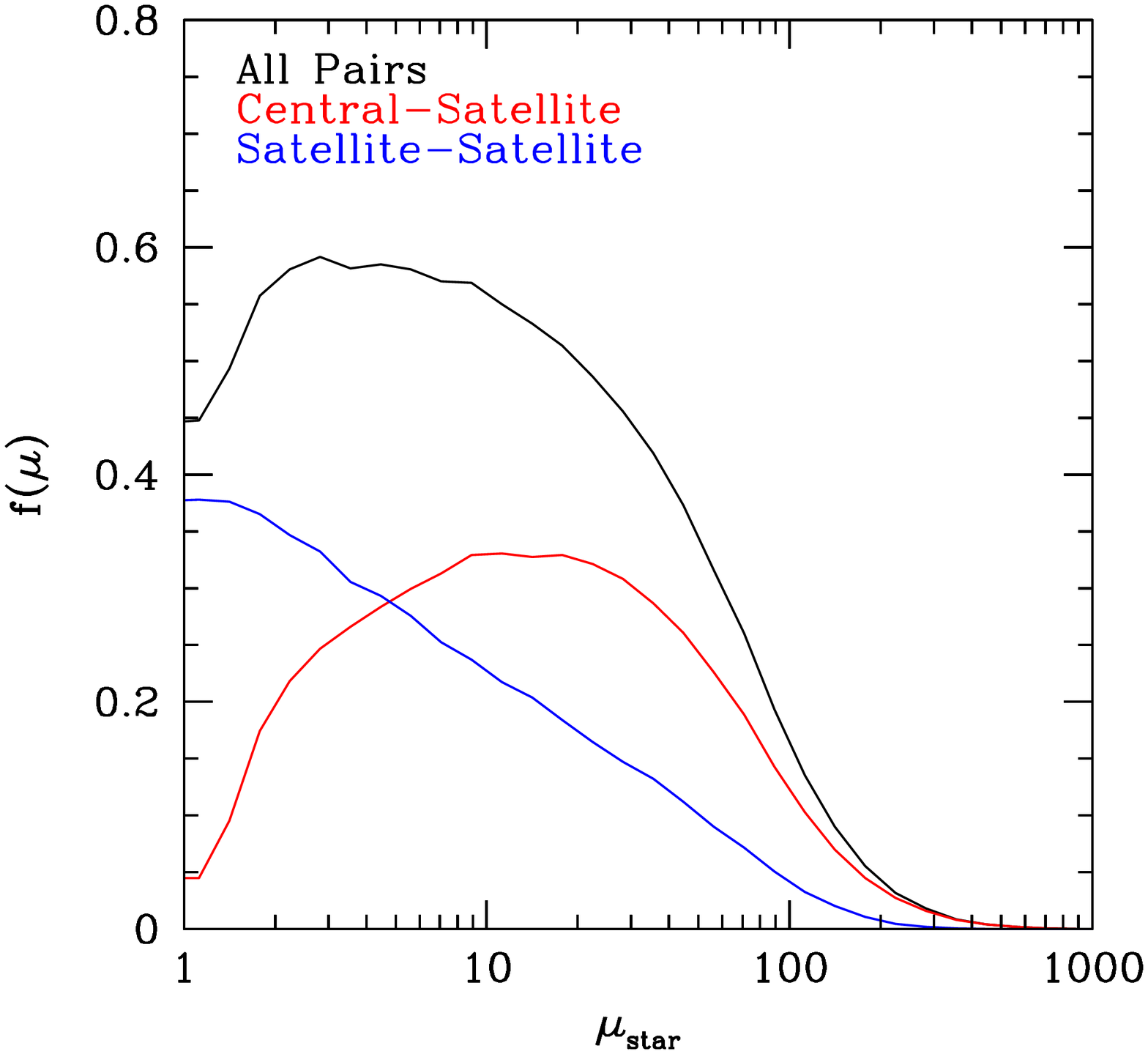}
 \caption{ 
{\it Left}: Distribution of stellar masses for individual galaxies in our main pair sample. Central-satellite pairs (red) tend to have intermediate-to-high stellar masses, peaking at m$_{\rm star}\sim10^{10.7}$M$_{\odot}$, and tail off at slightly below $10^{12}$M$_{\odot}$. Satellite-satellite pairs (blue), on the other hand, exhibit a more uniform distribution in stellar mass, and are able to probe (and slightly dominate) the low mass regime. However, these pairs never reach masses higher than $\sim2-3\times10^{11}$M$_{\odot}$. {\it Right}: Distribution of (stellar) mass ratios. By definition, $\mu_{\rm star}\geq 1$.  Central-satellite pairs tend to have intermediate mass ratios (peaking at $\mu_{\rm star}\sim$10-20), and can access extremely high mass ratios ($\mu_{\rm star}\gtrsim$ a few $\times 100$). However, these pairs tend to avoid the extremely major regime  ($\mu_{\rm star}\lesssim5$). Satellite-satellite pairs, however, tend to be slightly dominated by low-mass ratios ($\mu_{\rm star}\lesssim10$), and the numbers decrease monotonically with increasing mass ratio.}
 \label{figmmrf}
\end{figure*}

\subsubsection{Galaxy Pair Flavours: The Sample}\label{secflavsample}

For the sake of completeness, we also report the regions of stellar-mass/mass-ratio parameter space occupied by the two flavour subsets. Figure~\ref{figmmrf} is analogous to Figure~\ref{figsample}, but with contributions from the two flavours: central-satellite pairs (in red) and satellite-satellite pairs (in blue). The left panel shows the distribution of stellar masses of the individual galaxies in the main pair sample. At low masses, both flavours behave similarly, with the exception that the sample in that regime, the contribution from satellite-satellite pairs is slightly larger than that from central-satellite pairs. The opposite holds at the high-mass regime: central-satellite pairs tend to dominate there. Moreover, central-satellite pairs are capable of reaching the largest masses in the simulation.
Lastly, for the full sample (in black), the peak at $m_{\rm star} \sim 10^{10.7} M_{\odot}$ is primarily accounted for by a similar peak in the central-satellite distribution.

The right panel of Figure~\ref{figmmrf} shows the distribution of stellar mass ratios. The contribution from central-satellite pairs resembles that of the full sample, except that it peaks at slightly higher mass ratios, and precipitates faster at lower values. Also, the extreme high mass ratio regime is dominated by central-satellite pairs. As for the satellite-satellite pairs, the mass-ratio distribution decreases monotonically with increasing mass ratio. In this case, the low mass-ratio regime (containing pairs with nearly equal masses) is dominated by the satellite-satellite flavour. 

It is interesting to discuss our stellar mass-ratio distribution against what it is commonly found in observed close-pair samples. For instance, in the lower-right panel of their Figure 2, \cite{ellison10} show the distribution of stellar mass ratios from their SDSS close-pair catalogue (explicitly, they show five bins between $\log \mu_{\rm star}=0$ and 1). In that Figure, their distribution is strictly monotonically decreasing with increasing $\mu_{\rm star}$. In contrast, our main simulated sample (black solid curve, right-hand panel of Figure~\ref{figmmrf}) shows a dip below $\mu_{\rm star}\sim3$. Casting this distribution in terms of flavours sheds some light on this issue. Namely, this dip is explained entirely by the prominent dip below $\mu_{\rm star}\sim10$ present in the central-satellite mass-ratio distribution (solid red curve). Both dips are explained by the fact that, in the simulation, finding a pair of two nearly equal-mass interacting galaxies is very infrequent (in section~\ref{secEnergy} we demonstrate that the majority of central-satellite pairs are in fact bound systems). Therefore, we speculate that the lack of a dip in the observed $\mu_{\rm star}$ distribution is caused by projection effects, which may blur the signal. This issue, and several others, will be addressed in an upcoming paper, where we select our pair sample from a mock galaxy survey (Moreno et al., in prep). 

\subsection{The Third Neighbour}\label{secThird}

Our above analysis, based on halo membership, demonstrates the need to connect galaxy pairs to their surroundings. Namely, the location of the pairs relative to their corresponding dark matter halo(es) appears to be an important factor: central-satellite pairs behave differently from satellite-satellite pairs. However, from an observational standpoint, determining whether or not a given galaxy is a central or a satellite, or a member of a given dark matter halo, can be quite challenging \citep[e.g.,][]{yang07,skibba09,weinmann09}.  Inspired by Figure~\ref{figsloan}, in this section we explore the spatial location of our galaxy pairs within the cosmic web in a different way: by identifying the most influential massive galaxy in the surroundings.

\begin{figure*}
 \centering
    \includegraphics[width=\columnwidth]{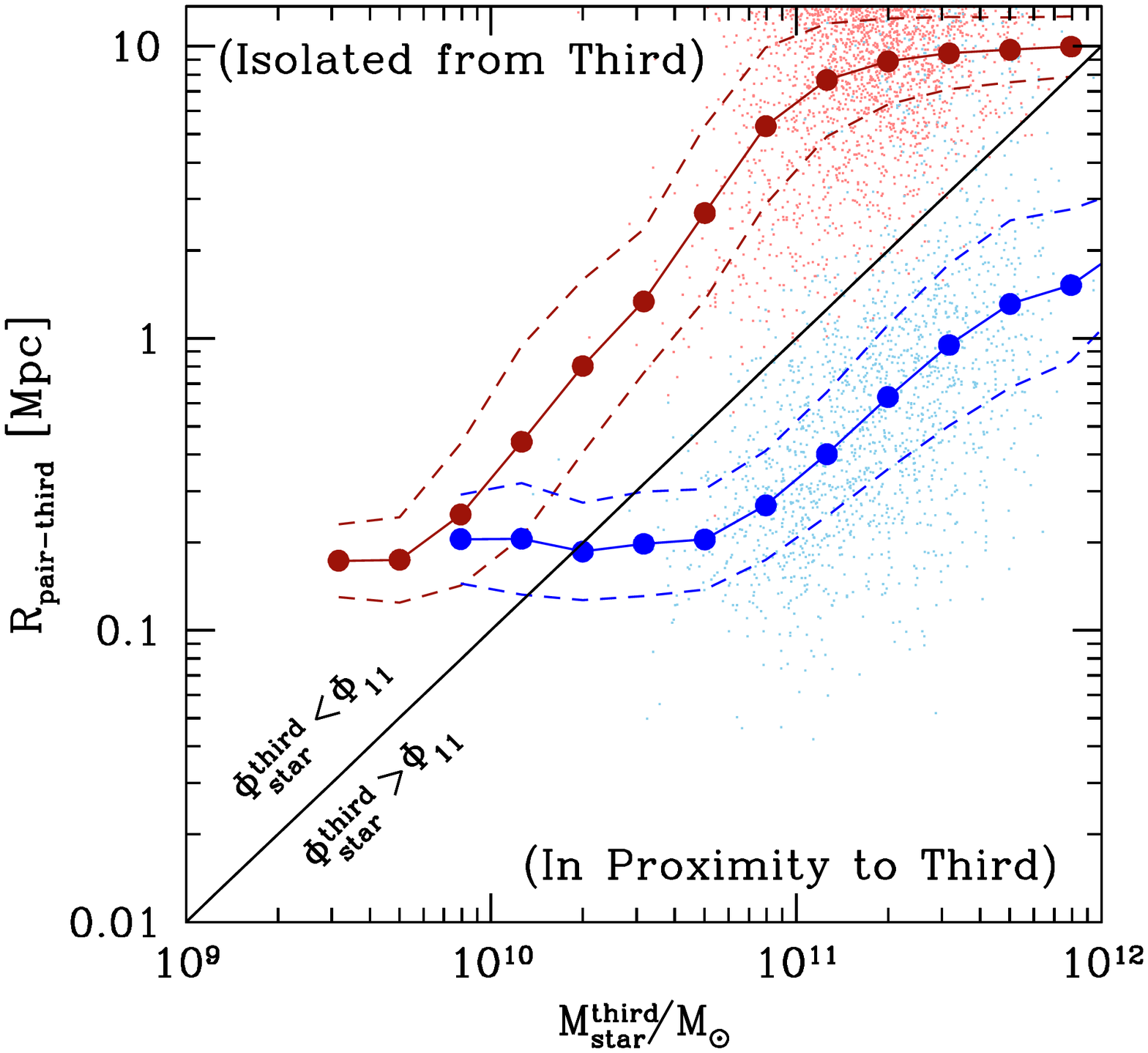}
  \includegraphics[width=\columnwidth]{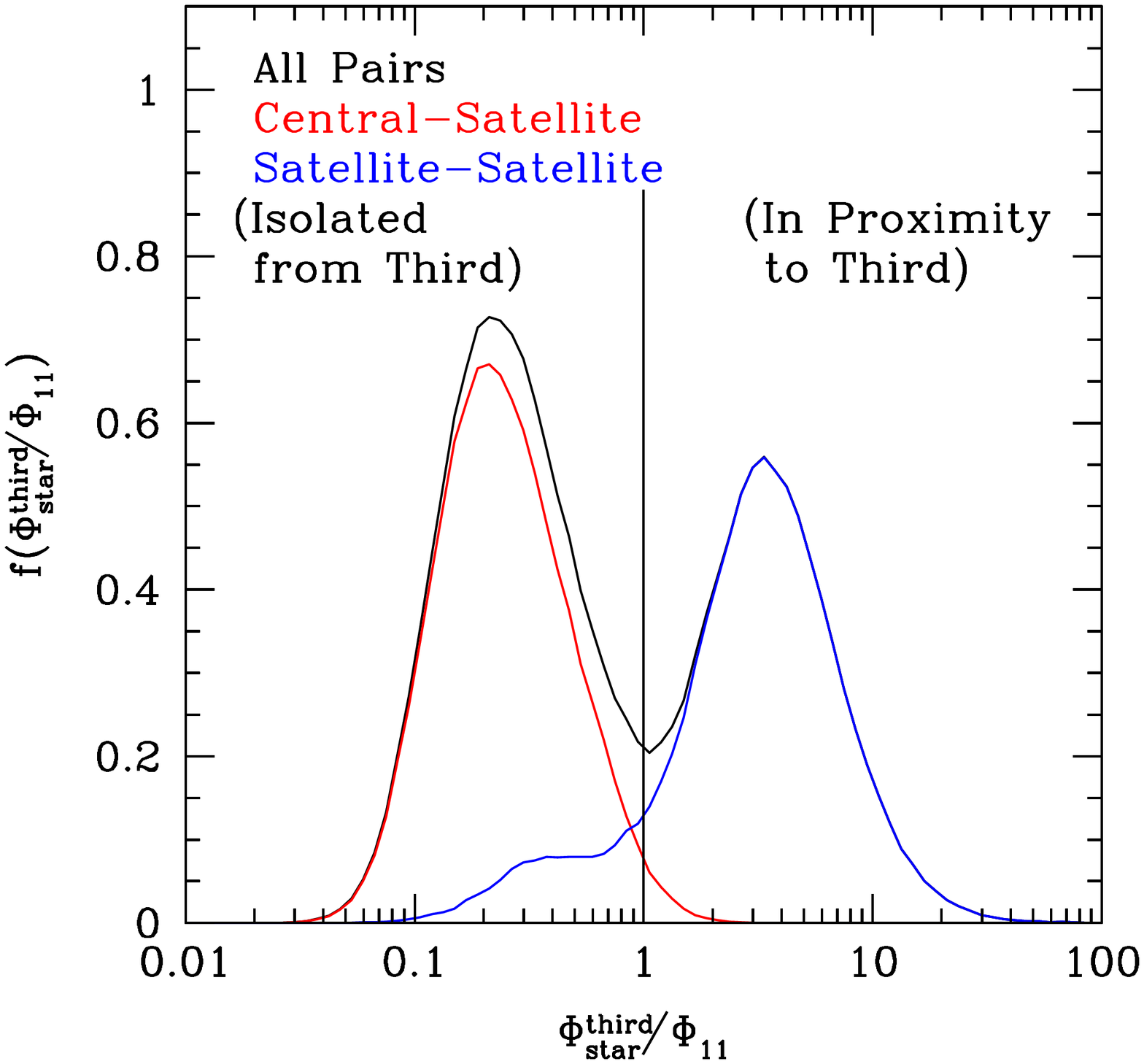}
 \caption{
 {\it Left}: Distance from the centre of mass of the pair to the third nearest/most-massive neighbouring galaxy (hereafter denoted by the label `third') versus the stellar mass of that galaxy. See equations~(\ref{eqnisodyn}) and (\ref{eqnisostar}), and associated text, for definitions. Diagonal solid line corresponds to a value of $\Phi^{\rm third}_{\rm star}$ fixed at 10$^{11}M_{\odot}/$Mpc $\equiv\Phi_{11}$.  This line roughly splits the pairs into two types: those isolated from the third, and those in close proximity to the third.  {\it Right}: Distribution of $\Phi^{\rm third}_{\rm star}$, the (stellar) proximity parameter. The black curve (all pairs) shows the bimodal nature of this distribution, with minimum at $\Phi^{\rm third}_{\rm star}\simeq\Phi_{11}$  Central-satellite pairs (red points and curves) tend to be in isolation, whilst satellite-satellite pairs (blue points and curves) prefer to inhabit regions in close proximity to their respective third neighbours.
 }
 \label{figiso}
\end{figure*}

Several methods attempting to quantify environment exist in the literature \citep{muldrew12,skibba13}, some of which are thought to be more directly connected to halo membership than others \citep{haas12}. It is beyond the scope of this paper to explore every possible environmental metric. Instead, motivated by our goal of investigating what drives the dynamics of galaxies in pairs, we propose the scheme described below. 

Consider a galaxy pair and all the galaxies surrounding them, irrespective of distance, only with the condition that said galaxies have {\it dynamical masses greater than those of either galaxy in the pair}. Next, we list all the neighbouring galaxies and rank them according to the following quantity:
\begin{equation}
\Phi^{\rm third}_{\rm dyn} \equiv \frac{{M^{\rm third}_{\rm dyn}}}{{R_{\rm pair-third}}},
\label{eqnisodyn}
\end{equation}
where  $M^{\rm third}_{\rm dyn}$ is the dynamical (dark matter) mass of the third neighbouring galaxy, and $R_{\rm pair-third}$ is the distance from the centre of mass of the pair to that object.  Hereafter, we call this quantity the (dynamical) {\bf proximity parameter}. Likewise, the galaxy in the vicinity with the {\it highest} value of this parameter is now called the {\bf third galaxy}. Notice that, with this definition, high values of $\Phi^{\rm third}_{\rm dyn}$ tend to select the most massive galaxies, but also penalize against very large separations. Also, we note that $R_{\rm pair-third}$ is {\it not} required to be greater than the separation between the two galaxies in the pair.

We note that in practice, we only search for neighbouring galaxies within a radius of 10 $h^{-1}$ Mpc from each pair. This results in $\sim$4\% of pairs {\it without} a third neighbour. We discuss the consequences of this approximation in Section~\ref{secDiscussion}. 

In forthcoming figures, we primarily employ a `stellar' version of the proximity parameter, which connects better to observations:
\begin{equation}
\Phi^{\rm third}_{\rm star} \equiv \frac{{M^{\rm third}_{\rm star}}}{{R_{\rm pair-third}}}.
\label{eqnisostar}
\end{equation}
Here, the mapping between the two proximity parameters is simply governed by the mapping between stellar and dynamical mass (equation~\ref{eqnmratio}).  

Figure~\ref{figiso} illustrates the types of (stellar) proximity parameters existing in main our galaxy pair sample. (Hereafter, we drop the word `stellar' and use the term `proximity parameter' to refer to $\Phi^{\rm third}_{\rm star}$, unless stated otherwise.) The left panel shows a scatter plot of distances from pairs to their respective third galaxies versus the stellar masses of those galaxies. In this figure, the value of our proximity parameter $\Phi^{\rm third}_{\rm star}$ increases from the upper-left corner (large separations and/or low-mass third galaxies) to the lower-right corner of the panel (small separations and/or massive third galaxies). 

The solid diagonal line, fixed at 10$^{11}M_{\odot}/$Mpc $\equiv\Phi_{11}$, roughly splits the panel into two sections: pairs in {\it close proximity} to their third massive neighbour (lower-right) and those {\it isolated} from their respective third galaxy (upper-left). The location of the line, whilst arbitrary, is empirically motivated by the distribution of proximity parameters (see below).  In this context, notice that our two dominant flavours are essentially segregated from each other:  central-satellite pairs (red points) tend to be in isolation relative to their third galaxies, whilst satellite-satellite pairs (blue points) prefer to be in close proximity to their third neighbours. It is also important to keep in mind that the degree of `proximity' is, in part, also governed by the mass of the third neighbour. For example, many central-satellite pairs are within only a few hundred $h^{-1}$ kpc of their respective third neighbours, but are considered isolated because those (third) galaxies have low masses.

\begin{figure*}
 \centering
  \includegraphics[width=\columnwidth]{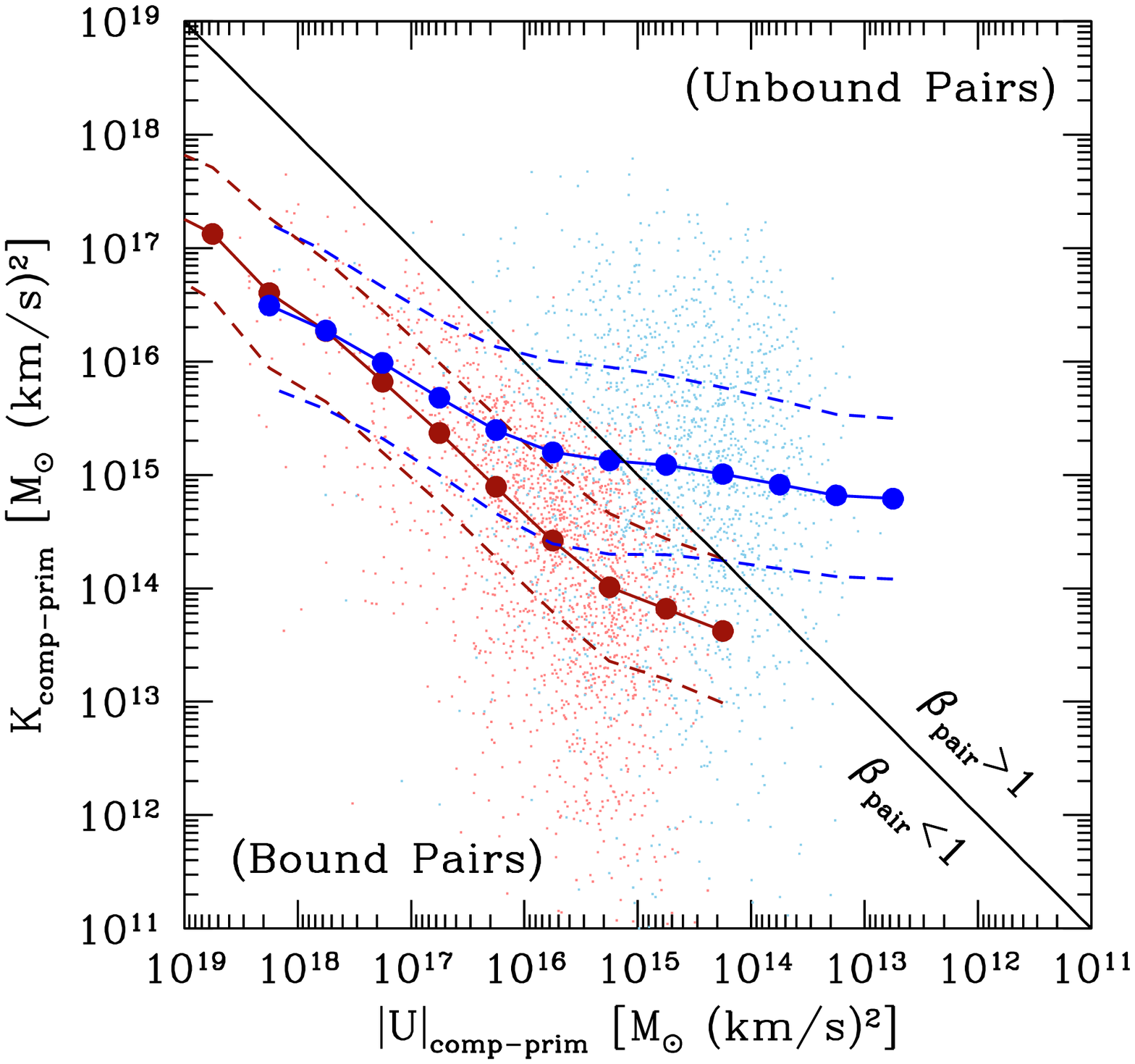}
  \includegraphics[width=\columnwidth]{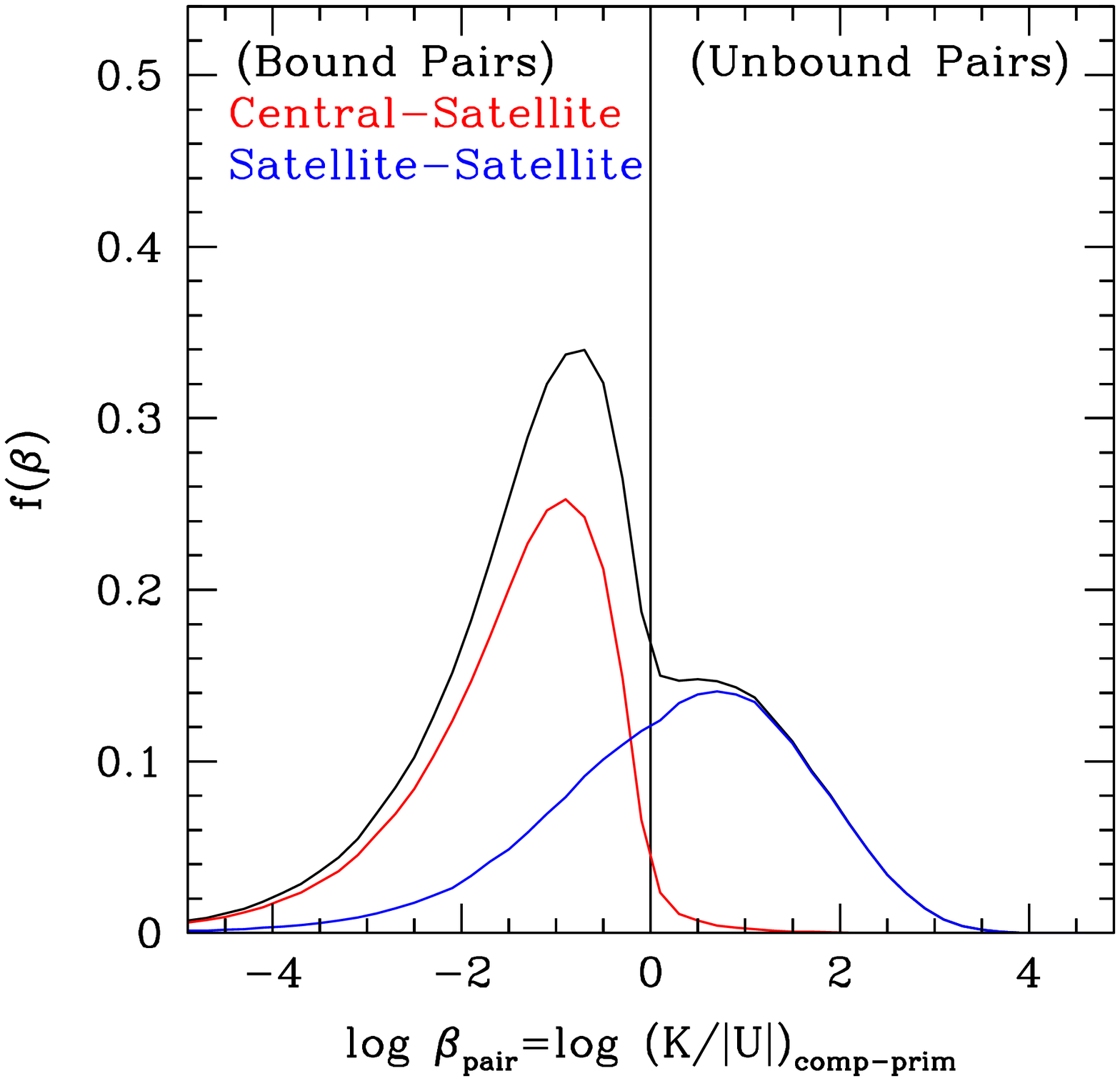}

 \caption{{\it Left}: Kinetic energy versus (the absolute value of) potential energy in galaxy pairs. The solid diagonal line denotes the zero-energy level. Pairs below and to the left of this line are {\it bound}, whist those above and to the right are {\it unbound}. {\it Right}: Distribution of binding parameters (equation~\ref{eqnbpair}). For the whole population (black curve), the distribution is rather bimodal. The notion of flavour offers a natural interpretation: central-satellite pairs (red curve) tend to be bound, whilst satellite-satellite pairs (blue curve) tend to be unbound, with the exception of a long tail of bound satellite-satellite pairs.}
 \label{figepair}
\end{figure*}

To illustrate, imagine the LMC-SMC pair (a satellite-satellite pair). Based on equations~(\ref{eqnisodyn}) and (\ref{eqnisostar}), the Milky Way would be the corresponding third galaxy and our pair would be considered to be in {\it close proximity} to it. Notice that in this particular example, the third galaxy is in the same dark matter halo as the pair itself. However, this is not required by construction, but most likely this is true for the majority of cases (whether or not this particular detail is true is largely irrelevant to our analysis). 

On the other hand, to illustrate the case of central-satellite pairs and their respective third neighbour, consider a group right outside a massive cluster. Here, imagine that the group's central galaxy makes a central-satellite pair with one of its satellites. In this case, the third galaxy would be the BCG at the centre of the massive nearby cluster. Notice that our pair and the BCG are in distinct dark matter haloes (the former is in a group-like halo, whilst the latter is in a cluster-like halo). This must be the case, by construction -- else if the pair and the BCG shared a halo, the pair would be categorized as a satellite-satellite pair, not a central-satellite pair. Given that the group containing the pair is outside the cluster (and, thus, at a large distance from the BCG) -- based on our equations~(\ref{eqnisodyn}) and (\ref{eqnisostar}), this pair would likely be classified as {\it isolated} from its third massive neighbour.

To give an additional example, imagine tracking the LMC-SMC pair back in time, to right before the moment this system was absorbed by the MW halo. In this case, this pair would also be treated as a central-satellite pair (in its own external halo), whilst the MW would take the role of the third galaxy, living in a separate dark matter halo.

Let us now quantify the nature of our proximity parameter (relative to a third massive neighbour) across our full galaxy pair sample. The right-hand panel of Figure~\ref{figiso} shows the probability distribution of $\Phi^{\rm third}_{\rm star}$ (modulo $\Phi_{11}$, indicated by the vertical black line), which is strongly bimodal: the majority of pairs ($\sim$57\%, left of the vertical line) are in the isolated regime (relative to their third neighbour), and the rest ($\sim$43\%, right of the vertical line) are in close proximity to a more massive object.  This histogram corroborates our previous findings (scatter plot to the left): with only a few exceptions, central-satellite galaxies are generally located far away from a third more massive galaxy, whilst satellite-satellite galaxies are not. This conclusion strengthens the central message of this paper: {\it treating pairs of galaxies (merging or not) as systems in pure isolation is merely an approximation}. The next section reiterates this point further, from an energy-based perspective.

\subsection{Binding Energy}\label{secEnergy}

\subsubsection{Binding Energy Within Pairs}\label{secenergypairs}

So far we have introduced two platforms to describe galaxy pairs and their cosmological context: halo membership/flavour (Figure~\ref{figflavs})  and degree of proximity to a third massive galaxy in the vicinity (equation~\ref{eqnisostar} and Figure~\ref{figiso}). A natural alternative is to use the binding energy holding the galaxies together (or not) in a pair.  At this point, one can ask the following questions: {\it Do central-satellite pairs tend to be energetically bound? Similarly, do satellite-satellite pairs tend to be unbound?}

To answer these questions, we introduce the notion of {\bf binding parameter of the pair}, defined as:
\begin{equation}
\beta_{\rm pair} \equiv (K/|U|)_{\rm comp-prim}.
\label{eqnbpair}
\end{equation} 
In this context, a pair is said to be bound if $\beta_{\rm pair}<1$, otherwise it is considered unbound. 
Here, $K$ represents the kinetic energy of the pair:
\begin{equation}
K_{\rm comp-prim}=\frac{1}{2}\mu^{\rm pair}_{\rm res,\,dyn}v^{2}_{\rm comp-prim},
\label{eqnkpair}
\end{equation}
where $v_{\rm comp-prim}$ is the velocity of the companion relative to the primary galaxy. We write this formula in terms of the residual\footnote{In this section, $\mu$ refers to residual mass, not mass ratio.} (dynamical) mass of the pair, defined as
\begin{equation}
\mu^{\rm pair}_{\rm res,\,dyn} \equiv \frac{m^{\rm prim}_{\rm dyn}m^{\rm comp}_{\rm dyn}}{m^{\rm prim}_{\rm dyn}+m^{\rm comp}_{\rm dyn}}.
\label{enqmpair}
\end{equation}
Similarly, $U$ represents the gravitational potential energy of the pair, given by 
\begin{equation}
U_{\rm comp-prim}\equiv-\frac{Gm^{\rm prim}_{\rm dyn}m^{\rm comp}_{\rm dyn}}{r_{\rm comp-prim}}+\frac{3}{10}(H\tilde{r})^2\times \Big(\frac{4}{3} \pi \rho_{\rm m} \tilde{r}^3\Big).
\label{eqnupair}
\end{equation}
Here, $r_{\rm comp-prim}$ is the separation between the galaxies in the pair, $\tilde{r}\equiv r_{\rm comp-prim}/2$, $H$ is the Hubble parameter, and $\rho_{\rm m}=\rho_{\rm crit}\times\Omega_{\rm m}$. The first term represents the standard Newtonian gravitational potential, whilst the second term accounts for the Hubble expansion of the Universe \citep{tonnesen12}.

With these definitions in place, we now present a scatter plot showing the kinetic energy versus the absolute value of the potential energy ($K$ versus $|U|$, left-panel of Figure~\ref{figepair}). We adopt this choice of axes to highlight the fact when $K$ is greater that $|U|$, the two galaxies in a pair are unbound (i.e., $\beta_{\rm pair}=K/|U|>1$) -- whilst the opposite holds when $|U|$ is greater than $K$ (i.e., the pair is a bound system, and $\beta_{\rm pair}=K/|U|<1$). Thus, in this figure, our binding parameter, $\beta_{\rm pair}$, increases from the lower-left corner  (with the `Bound Pairs' label) to the upper-right corner of the figure (with the `Unbound Pairs' label). The diagonal line at $\beta_{\rm pair}=1$ (i.e., with $K=|U|$) demarcates the boundary that separates bound from unbound pairs. 

This figure immediately shows that {\it central-satellite pairs tend to be bound} (red points below the diagonal), whilst {\it the majority of satellite-satellite pairs are unbound} (blue points above the diagonal). However, further inspection also shows that {\it a significant fraction of satellite-satellite pairs are energetically bound} (blue points below the diagonal). 

\begin{figure*}
 \centering
 \includegraphics[width=\columnwidth]{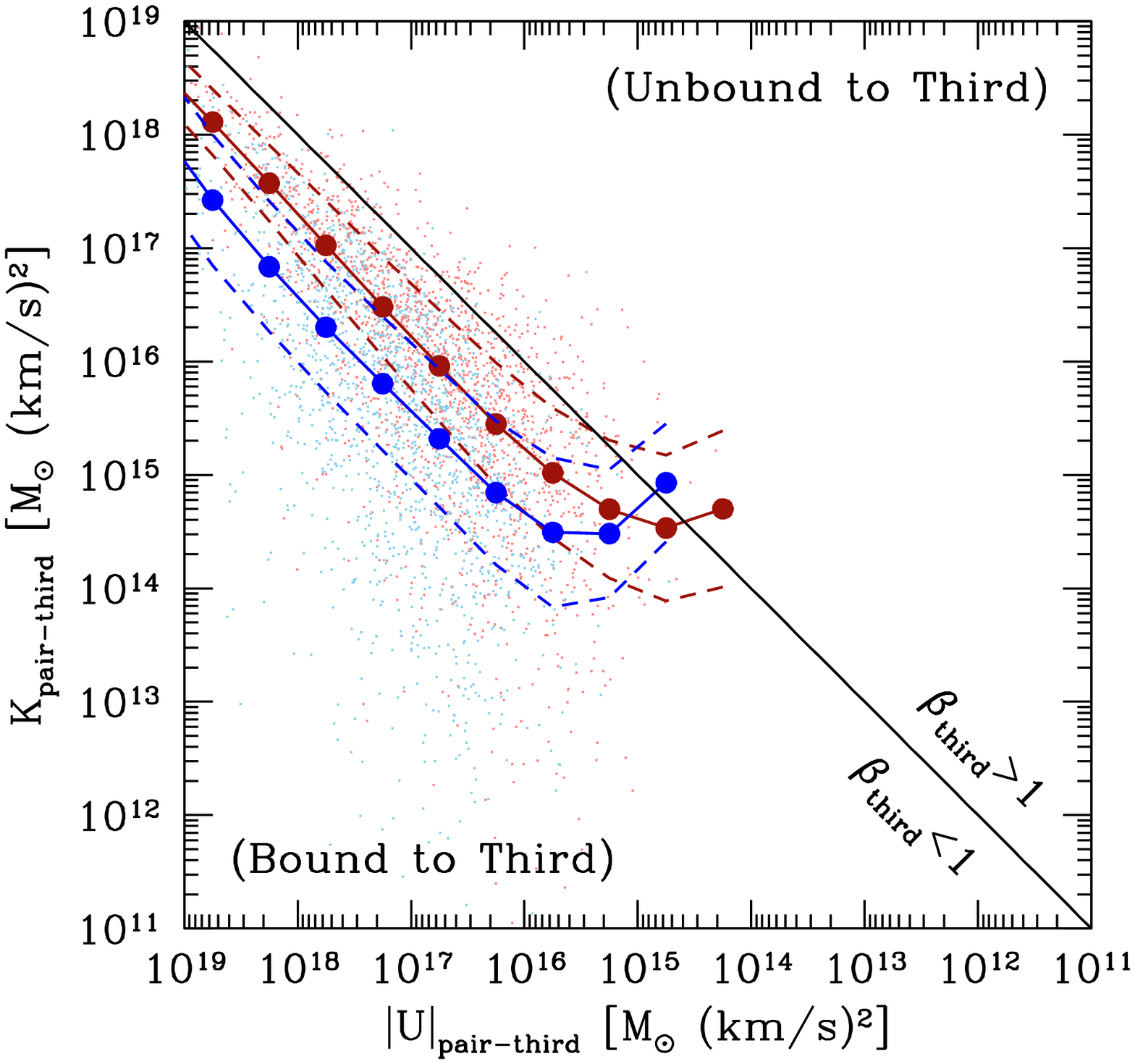}
  \includegraphics[width=\columnwidth]{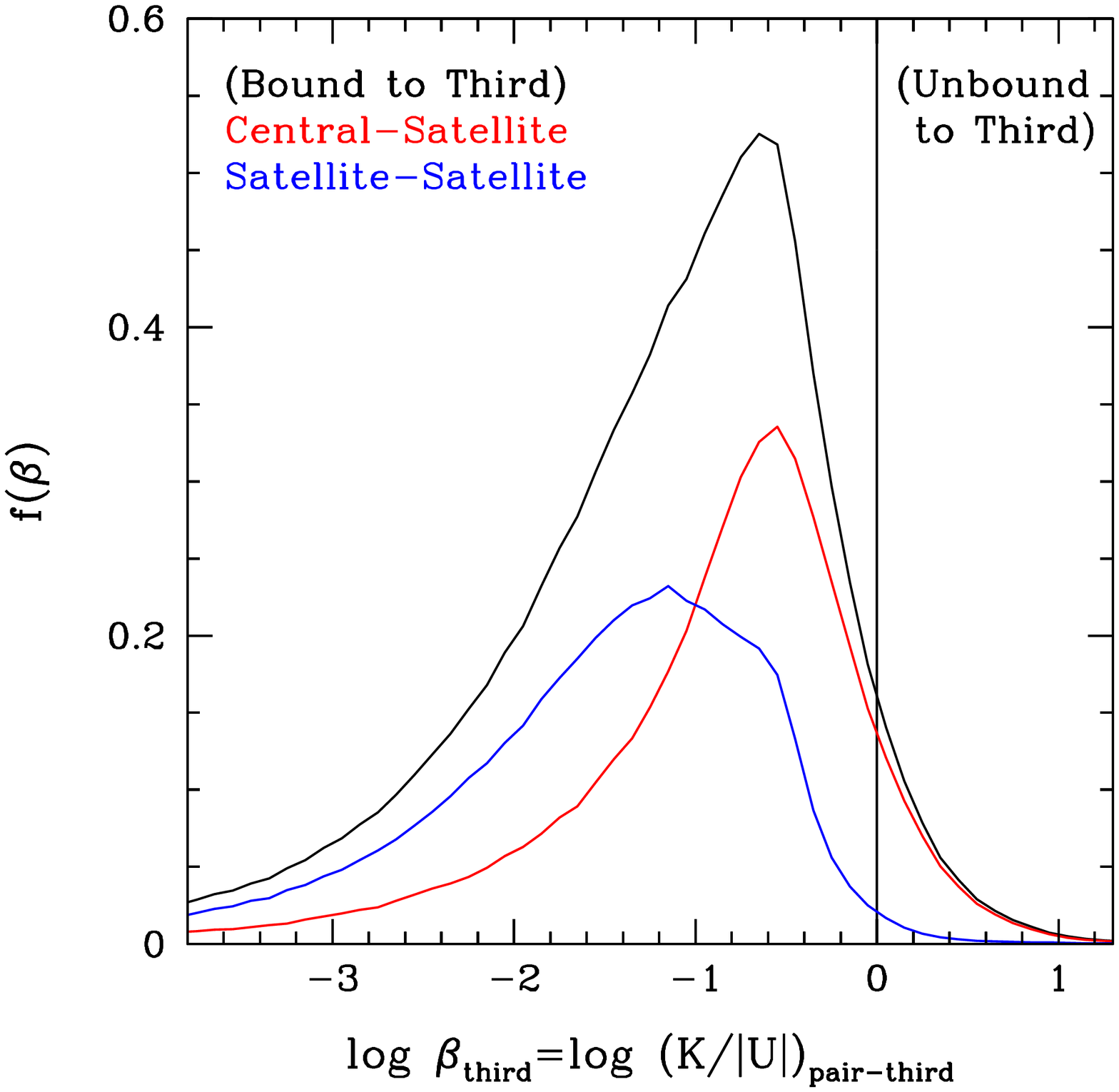}
\caption{{\it Left}: Kinetic energy versus (the absolute value of) potential energy in pair-third systems. The solid diagonal line denotes the zero-energy level. Pairs below and to the left of this line are {\it bound} to their third nearest-massive neighbouring galaxy, whist those above and to the right are {\it not}. {\it Right}: Distribution of binding parameters (equation~\ref{eqnbthird}) of pairs relative to the third. For the whole population (black curve), the distribution is peaks well within the `bound-to-the-third' regime, with 77.3\% of all pairs being bound to a third massive galaxy. Satellite-satellite pairs (blue curve) tend to be more strongly bound to a third galaxy than central-satellite pairs (red curve), although the majority of central-satellite pairs are in fact bound to a third galaxy as well. The population of pairs unbound to a third galaxy is dominated by central-satellite pairs.}
 \label{figethird}
\end{figure*}

These trends are quantified in the right-panel of Figure~\ref{figepair}, which shows histograms of $\beta_{\rm pair}$, the binding parameter. This figure is split by a vertical line at $\beta_{\rm pair}=1$. The distribution for the whole population is bimodal: $\sim$70\% of the pairs are bound (left of the vertical line), whilst $\sim$30\% are unbound (right of the vertical line). In terms of flavours, one can readily see that the great majority of central-satellite pairs tend to be bound ($\sim$98\%). However, whilst the majority of satellite-satellite pairs ($\sim$62\%) are unbound (as expected), a large fraction of them ($\sim$38\%) actually count as bound pairs. 

To summarize, the overwhelming majority of central-satellite pairs are bound, whilst a marginal majority of satellite-satellite pairs (slightly over 60\%) are unbound. However, almost 40\% of satellite-satellite pairs are actually bound systems. This is rather intriguing, and suggests that {\it a significant fraction of pairs of satellites are in fact energetically bound to each other, and possibly in the process of merging}. For example, this may be the case for the satellite-satellite pairs found in massive galaxy clusters \citep[][]{vandokkum99,tran05}. 

\subsubsection{Binding Energy Relative to the Third Neighbour}\label{secenergypairs}

We devote this section to quantifying the binding energy of the pairs relative to their third massive neighbour (see equation~\ref{eqnisodyn} and accompanying text).
As in the previous section, we ask the following questions: {\it Do satellite-satellite pairs tend to be bound to their third neighbour? Similarly, do central-satellite pairs tend to be unbound to their corresponding third massive galaxies in the vicinity?}

To answer these questions, we define the {\bf binding parameter of the pair-third system}:
\begin{equation}
\beta_{\rm third} \equiv (K/|U|)_{\rm pair-third}.
\label{eqnbthird}
\end{equation}
This parameter allows us to distinguish pairs that are energetically bound to their respective third galaxy ($\beta_{\rm pair-third}<1$) from those that are not ($\beta_{\rm pair-third}>1$). In this equation, $K$ denotes the kinetic energy of the pair relative to third, and is given by 
\begin{equation}
K_{\rm pair-third}=\frac{1}{2}\mu^{\rm third}_{\rm res,\,dyn}V^{2}_{\rm pair-third},
\label{eqnkthird}
\end{equation}
where $V_{\rm pair-third}$ is the magnitude of the vectorial difference between the centre-of-mass velocity of the pair and the velocity of the third. For reference, the centre-of-mass velocity is given by
\begin{equation}
\vec{V}^{\rm pair}_{\rm C.M.} \equiv \frac{m^{\rm prim}_{\rm dyn}\vec{v}_{\rm prim}+m^{\rm comp}_{\rm dyn}\vec{v}_{\rm comp}}{m^{\rm prim}_{\rm dyn}+m^{\rm comp}_{\rm dyn}}.
\label{eqnvcm}
\end{equation}
Also, in equation~(\ref{eqnkthird}), the residual (dynamical) mass associated with the pair-third system is defined as
\begin{equation}
\mu^{\rm third}_{\rm res,\,dyn} \equiv \frac{M^{\rm pair}_{\rm dyn}M^{\rm third}_{\rm dyn}}{M^{\rm pair}_{\rm dyn}+M^{\rm third}_{\rm dyn}},
\label{enqmpair}
\end{equation}
where $M^{\rm pair}_{\rm dyn}\equiv m^{\rm prim}_{\rm dyn}+m^{\rm comp}_{\rm dyn}$.

Similarly, $U$ refers to the potential energy of the pair relative to the third, and is given by 
\begin{equation}
U_{\rm pair-third}\equiv-\frac{GM^{\rm pair}_{\rm dyn}M^{\rm third}_{\rm dyn}}{R_{\rm pair-third}}+\frac{3}{10}(H\tilde{R})^2\times \Big(\frac{4}{3} \pi \rho_{\rm m} \tilde{R}^3 \Big),
\label{eqnuthird}
\end{equation}
where $R_{\rm pair-third}$ is the separation between the centre-of-mass of the pair and the third, and $\tilde{R}\equiv R_{\rm pair-third}/2$. Compare equations~(\ref{eqnkthird}) and (\ref{eqnuthird}) to equations~(\ref{eqnkpair}) and (\ref{eqnupair}).

Before discussing our results, we highlight that are using {\it dynamical} (as opposed to stellar) masses in our calculations. For instance, if the third galaxy associated to a given pair corresponds to a BCG, the appropriate mass to use is that of its host (cluster-like) dark matter halo. In the discussion that follows, we shall casually use the phrase `binding energy relative to the third galaxy' -- by which we strictly mean `binding energy relative to the dark matter halo/subhalo hosting the third galaxy'. Nevertheless, the potentially stronger connection to observations motivates us to continue to speak of energy relative to the third neighbouring {\it galaxy}, without direct reference to its dark host.

Figure~\ref{figethird} is analogous to Figure~\ref{figepair}, but now focusing on the binding parameter corresponding to the pair-third system. The left panel of Figure~\ref{figethird} shows a scatter plot of the kinetic energy versus the absolute value of the potential energy. The diagonal line at $\beta_{\rm third}=1$ separates those pairs bound to a third galaxy (lower-left half) from those that are {\it not}  (upper-right half). The first thing to notice is that {\it satellite-satellite pairs tend to be more strongly bound to a third more-massive galaxy than their central-satellite counterpart}s (blue versus red points). However, this is {\it not} equivalent to saying that the $\beta_{\rm third}=1$ line splits the sample evenly into two clearly discernible sub-populations (i.e., satellite-satellite pairs from central-satellite pairs, as in Figure~\ref{figiso}). Instead, the figure suggests that, in fact, {\it the great majority of central-satellite pairs are also bound to a third more-massive neighbouring galaxy}. 

The histograms displayed in the right-panel of Figure~\ref{figethird} quantify these trends. First note that essentially all satellite-satellite pairs (almost 99\% of them) are bound to a third neighbouring galaxy (the portion of the blue curve lying to the left of the vertical line). Surprisingly, the great majority of central-satellite pairs ($\sim$83\%) are also bound to a third galaxy (the portion of the red curve lying to the left of the vertical line). In fact, {\it the majority of galaxy pairs in our sample (i.e., $\sim$91\%) are actually also bound to a third massive object in the vicinity} (the portion of black curve to the left of the vertical line). 

To summarize, it is interesting to ask the following question: {\it what fraction of our cosmological sample most closely resembles the kind of galaxies modelled by idealised binary merger simulations?} (E.g., the sort of suites of simulations by Di Matteo et al. 2007, Cox et al. 2008, and Torrey et al. 2012, to name a few.) Our results indicate that, indeed, only a small minority of pairs are energetically isolated. In other words, {\it only $\sim$9\% of the pairs can be regarded as evolving in pure isolation, mirroring the situations described by hydrodynamical merger simulations}. (Explicitly, this fraction includes $\sim$5\% of those that have a third galaxy identified within a sphere 10 $h^{-1}$ Mpc, plus $\sim$4\% without such a neighbour in a sphere of that size.)
This conclusion is quite remarkable, and resonates well with the main message of our paper. Namely, {\it the influence of environment is inevitable, and plays a critical role in determining the dynamical evolution of most galaxy pairs!}

\begin{figure*}
 \centering
  \includegraphics[width=\columnwidth]{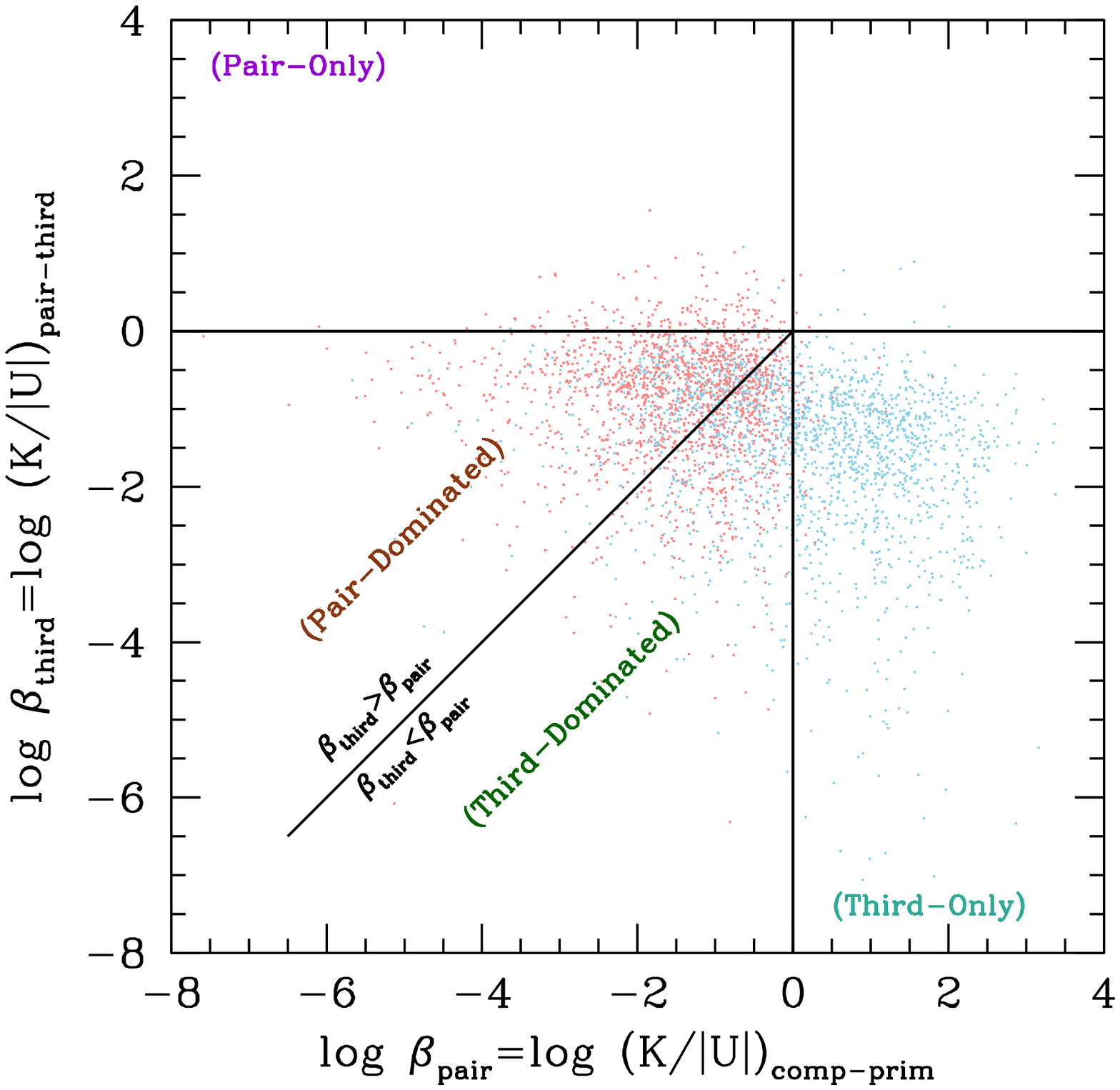}
    \includegraphics[width=\columnwidth]{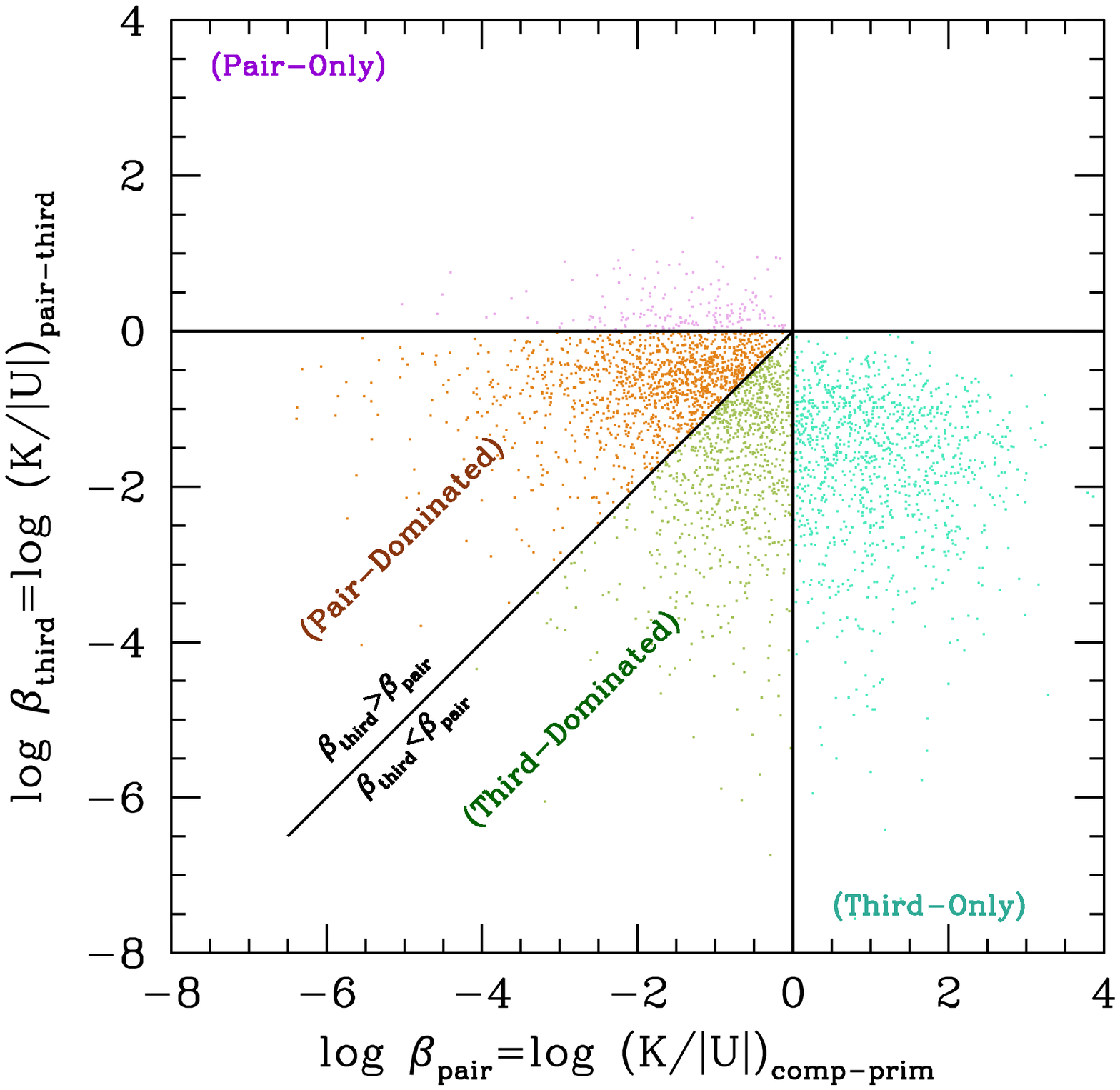}

 \caption{{\it Left}: Binding parameter of pair-third systems versus binding parameter of the pairs ($\beta_{\rm third}$ versus $\beta_{\rm pair}$ -- see equations~\ref{eqnbpair} and \ref{eqnbthird} for definitions).  As in previous figures, red points correspond to central-satellite pairs, and blue points represent satellite-satellite pairs. The vertical line splits the sample into bound and unbound pairs, whilst the horizontal line splits it into pairs bound and unbound to a third massive galaxy. Pairs in the upper-left quadrant are only bound to themselves, whilst those in the lower-right quadrant are unbound, yet bound to a third massive galaxy. Pairs in the lower-left quadrant are both self-bound, and bound to a third massive galaxy. The diagonal splits this quadrant into two regions: one where the pairs dominate (above the diagonal), and one where a third massive galaxy dominates (below the diagonal). {\it Right}: Identical to the left-panel, but in terms of our energy classes: {\it Pair-Only} ($\beta_{\rm pair}<1,\beta_{\rm third}>1$ -- PO, violet points), {\it Pair-Dominated} ($\beta_{\rm pair}<1,\beta_{\rm third}<1$, $\beta_{\rm pair}<\beta_{\rm third}$ -- PD, brown points), {\it Third-Dominated} ($\beta_{\rm pair}<1,\beta_{\rm third}<1$, $\beta_{\rm pair}>\beta_{\rm third}$ -- TD, green points), and {\it Third-Only} ($\beta_{\rm pair}>1,\beta_{\rm third}<1$ -- TO, turquoise points). Hereafter, the spurious pairs in the upper-right corner (neither bound to themselves nor to a third galaxy) are discarded.}

 \label{figbb}
\end{figure*}

\subsection{A Comparison: Energy Binding Within Pairs versus Binding to the Third Neighbour}\label{secPairThird}

Our results show that, in general, one cannot assume that pairs of galaxies evolve in isolation. In particular, we find that the majority of pairs in our sample are bound to a third neighbour. However, it is also true that many of those pairs are themselves self-bound entities. In other words, our numbers implicitly reveal that a large portion of the galaxy pairs are {\it both} self-bound {\it and} bound to a third galaxy.

For this reason, it is important to compare the degree of binding within pairs to how bound these are to a third object. Figure~\ref{figbb} illustrates this comparison. In the left panel, we show the binding parameter within the pair ($\beta_{\rm pair}$ -- equation~\ref{eqnbpair}) versus the binding parameter of the pair-third system ($\beta_{\rm third}$ -- equation~\ref{eqnbthird}). Points to the left of the vertical line are bound ($\beta_{\rm pair}<1$), whilst those to the right are unbound ($\beta_{\rm pair}>1$). Similarly, points below the horizontal line are bound to a third neighbouring galaxy ($\beta_{\rm third}<1$), whilst those above it are not ($\beta_{\rm third}>1$).

In terms of galaxy pair flavours (Section~\ref{secFlavs}), this figure shows that the majority of pairs that are self-bound, and unbound to a third massive galaxy, tend to be central-satellite pairs (red points in the upper-left quadrant). At the other extreme, the majority of pairs that are unbound systems, and bound to a third galaxy, tend to be satellite-satellite pairs (blue points in the lower-right quadrant). The lower-left quadrant, which contains pairs that are both self-bound and bound to a third galaxy, comprises a mixture of central-satellite and satellite-satellite pairs that cannot be easily disentangled. In fact, the majority of pairs in our sample lie in this quadrant; namely, $\sim$60\% of our pairs are both self-bound, and bound to a third neighbouring galaxy! 

Lastly, we draw a diagonal line in that quadrant, whose purpose is to separate pairs where self-binding dominates (above the diagonal, with $\beta_{\rm pair}<\beta_{\rm third}$) from pairs where binding to a third neighbour dominates (below the diagonal, with $\beta_{\rm pair}>\beta_{\rm third}$). (The upper-right quadrant contains spurious pairs that are not self-bound, nor bound to a third neighbour. These amount to under 0.8\%, and will be ignored for the rest of the paper.) In the next section, we explore the consequences of classifying our galaxy pairs in terms of binding energy.

\subsection{Energy Classes: Basic Definitions}\label{secClasses}

The previous section demonstrates that the majority of galaxies in pairs are bound to their companions {\it and} to a third massive galaxy in the vicinity. The rest are either bound pairs that are not bound to a third galaxy -- or simply the exact opposite: unbound pairs that happen to be bound to a third neighbour. In this section, we compare the degree of binding within the pairs versus the binding energy relative to a third massive neighbour.

The right panel of Figure~\ref{figbb} now colour-codes points according to the quadrant they occupy in the $\beta_{\rm pair}-\beta_{\rm third}$ plane. With this picture in mind, we postulate the following {\it four} {\bf energy classes}\footnote{In the next few sections, we encourage the reader to pay attention to the following abbreviations: {\bf PO}, {\bf PD}, {\bf TD} and {\bf TO}.}:

\begin{itemize}
\item {\bf Pair-Only} ({\it PO} -- 9.6\%): bound pairs that are not bound to a third massive galaxy ($\beta_{\rm pair}<1$, $\beta_{\rm third}>1$ -- violet points, upper-left quadrant).  \\

\item {\bf Pair-Dominated} ({\it PD} -- 35.4\%): bound pairs that are also bound to a third massive galaxy, but the pair dominates ($\beta_{\rm pair}<1$, $\beta_{\rm third}<1$, $\beta_{\rm pair}<\beta_{\rm third}$ -- brown points, lower-left quadrant above the diagonal).\\

\item {\bf Third-Dominated} ({\it TD} -- 25.4\%): bound pairs that are also bound to a third massive galaxy, but the third dominates ($\beta_{\rm pair}<1$, $\beta_{\rm third}<1$, $\beta_{\rm pair}>\beta_{\rm third}>1$ -- green points, lower-left quadrant below the diagonal).\\

\item {\bf Third-Only} ({\it TO} -- 29.3\%): unbound pairs that are bound to a third massive galaxy, ($\beta_{\rm pair}>1$, $\beta_{\rm third}<1$ -- turquoise points, lower-right quadrant).\\

\end{itemize}

See also the fifth column Table~\ref{tablecumd}, which explicitly quotes the relative contributions of each energy class to our full galaxy pair sample. (The other columns in that table quote similar percentages, but instead for subsamples with more stringent separation criteria, which are discussed in Section~\ref{seccumd}.) Also recall that spurious pairs not bound to each other nor to a third galaxy are no longer discussed, and are absent in right-panel of Figure~\ref{figbb} (i.e, the upper-right quadrant is now empty).

\begin{figure*}
 \centering
  \includegraphics[width=\columnwidth]{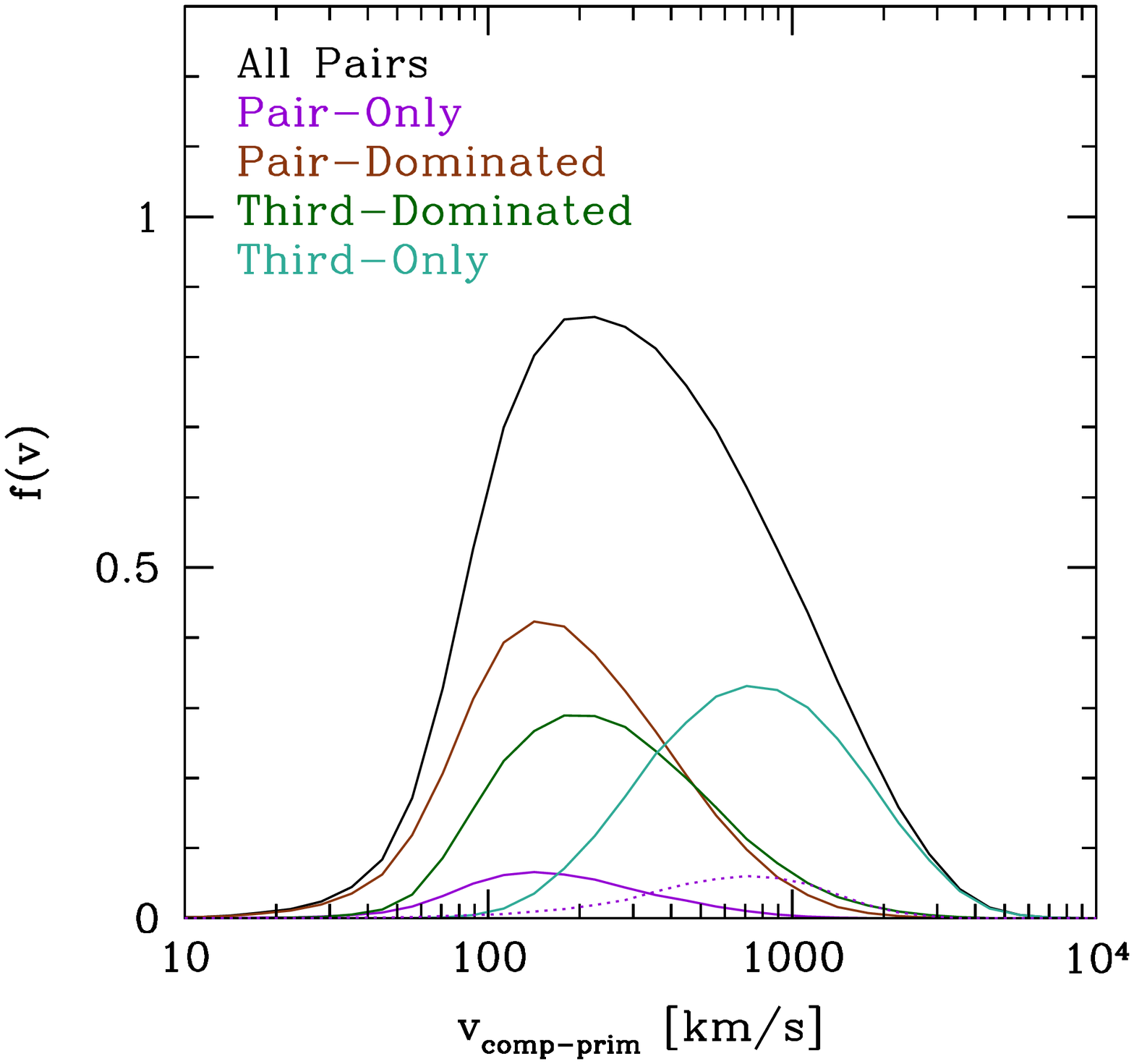}
  \includegraphics[width=\columnwidth]{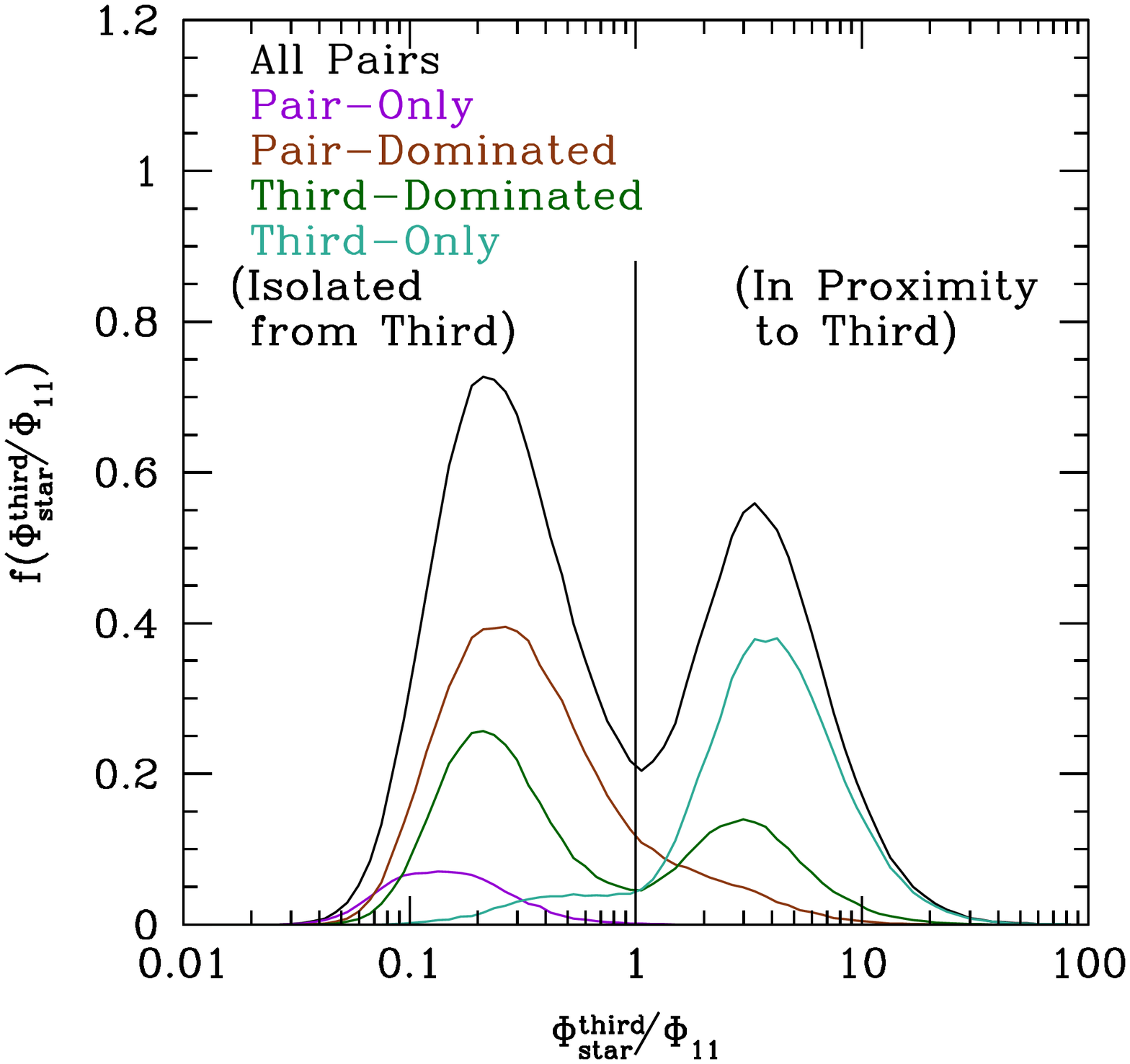}
 \caption{
 Distribution of relative velocities ({\it left}) and proximity parameter ({\it right} -- see equation~\ref{eqnisostar} for definition), in terms of our four energy classes: Pair-Only (PO -- violet curves), Pair-Dominated (PD -- brown curves), Third-Dominated (TD -- green curves), and Third-Only (TO -- turquoise curves). Bound pairs with no third neighbour within 10 $h^{-1}$ Mpc are also included (hereafter labelled IsoPO -- dotted violet curves). See Figure~\ref{figbb} and associated text for definitions. The first three energy classes (PO, PD and TD) exhibit similar relative velocities (peaking at $\sim$ 100-200 km sec$^{-1}$) and tend to be isolated from a third massive galaxy (although the TD class has a smaller peak in the high proximity-parameter regime). Pairs in the last energy class, TO, behave differently: they tend to approach higher relative velocities (approaching $\sim$1,000 km sec$^{-1}$) and are typically in isolation relative to the third. Lastly, those pairs with no third neighbour identified within 10 $h^{-1}$ Mpc (labelled IsoPO) also tend to have high relative velocities. By definition, the proximity parameter is not defined for IsoPO pairs (hence the lack of a dotted violet curve in the right-hand panel).
} 
 \label{figqphys}
\end{figure*}

First, let us discuss the two `extreme' energy cases: the Pair-Only class and the Third-Only class (respectively, upper-left and lower-right quadrants in Figure~\ref{figbb}). The first class (PO -- violet points) refers to pairs involving two galaxies that are {\it only} bound to one another. From an energetic point of view, galaxies in this class closely resemble those considered in idealised binary merger simulations in isolation. The last class (TO -- turquoise points), on the other hand, refers to pairs involving two galaxies that are {\it only} bound to a third more massive neighbour, and not to each other. These are merely two galaxies orbiting a common halo that happen to be in close proximity by mere chance.

We now comment on the two `intermediate' energy classes: the Pair-Dominated class and the Third-Dominated class (the lower-left quadrant in Figure~\ref{figbb}). First note that the pairs classified as PD (brown points) are very similar to those classified as PO (violet points). The only difference is that the former are bound to a third galaxy, whilst the latter are not. However, in the case of PD, the strength of how bound the primary and the companion galaxies are to each other still overwhelms the binding of this pair to their corresponding third neighbour. On the other hand, pairs classified as TD behave rather differently: whilst these galaxies are still bound to each other, their energetic connection is weak compared to the binding relative to a third massive neighbour in the vicinity. 
 
\subsection{Energy Classes: Analysis}

In the last section we provided a description in terms of two binding parameters, $\beta_{\rm pair}$ and $\beta_{\rm third}$ -- which allow us to split our main galaxy-pair sample into four energy classes: PO, PD, TD and TO. In this section, we revisit the results discussed in Sections~\ref{secFlavs} and \ref{secThird}, but now cast in terms of energy classes.

\begin{figure*}
 \centering
   \includegraphics[width=\columnwidth]{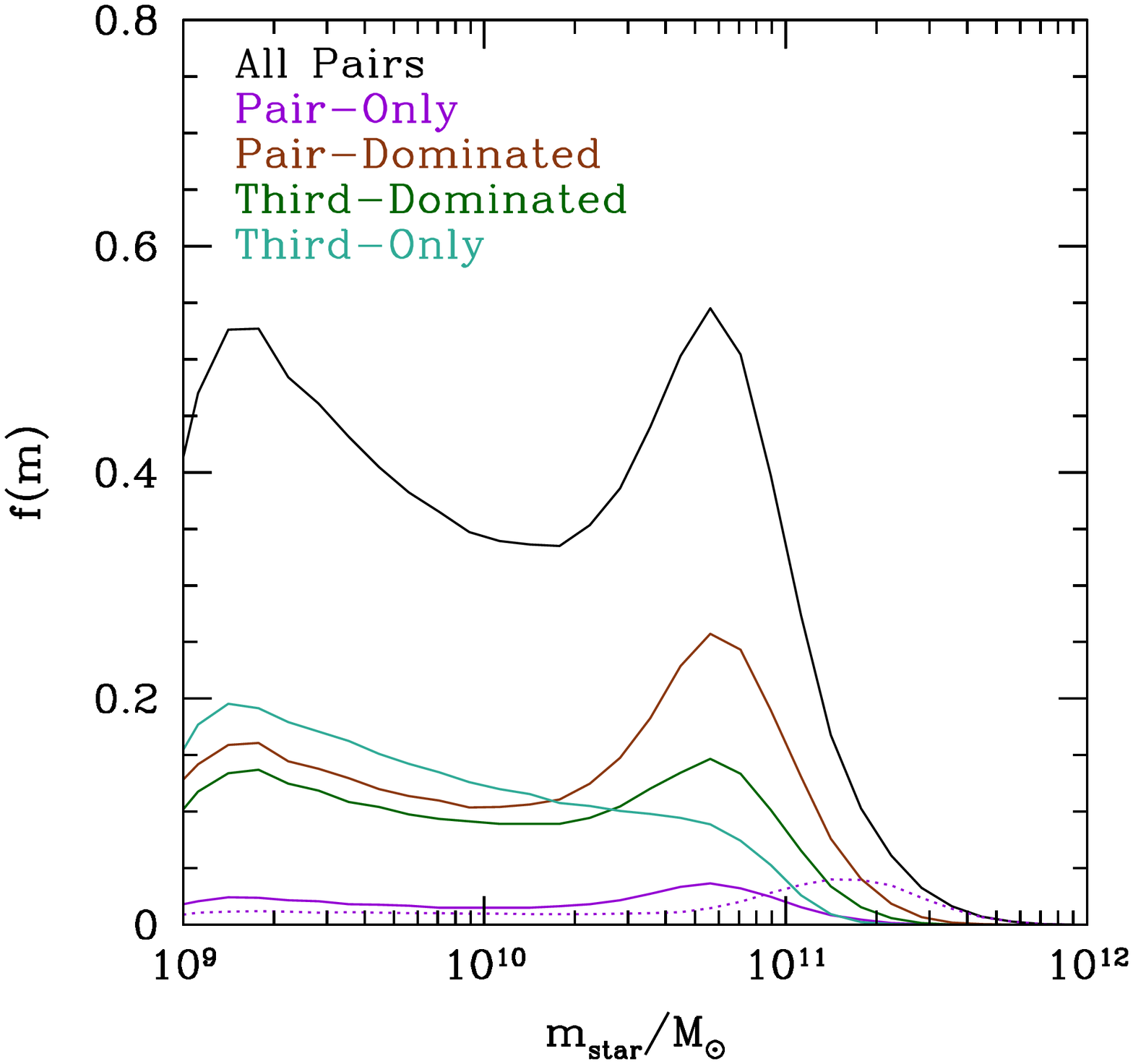}
  \includegraphics[width=\columnwidth]{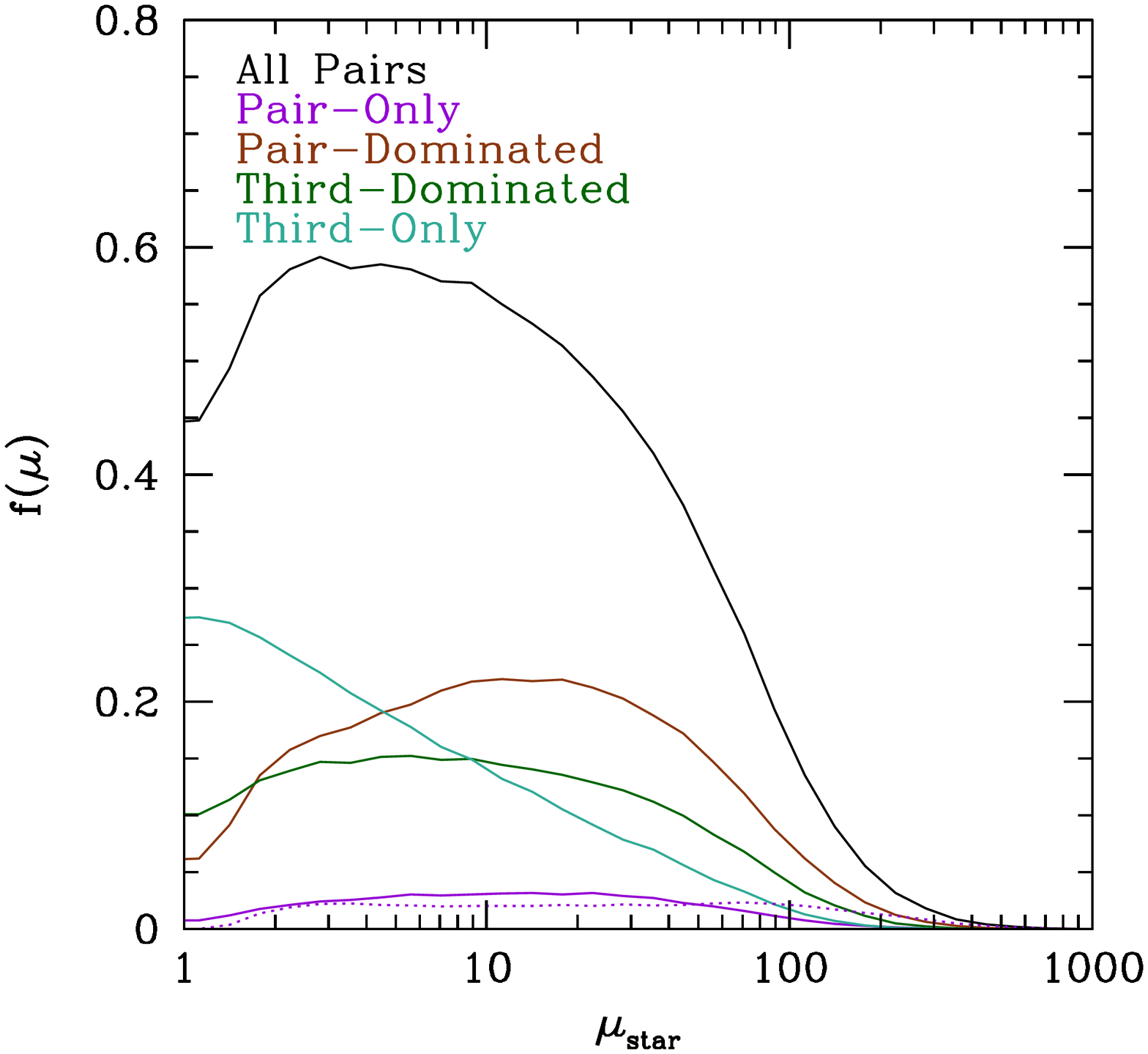}
 \caption{Distribution of individual stellar masses ({\it left}) and stellar-mass ratios ({\it right}) in terms the four energy classes: Pair-Only (PO -- violet curves), Pair-Dominated (PD -- brown curves), Third-Dominated (TD -- green curves), and Third-Only (TO -- turquoise curves). Bound pairs with no third neighbour within 10 $h^{-1}$ Mpc are also included (IsoPO -- dotted violet curves). The first three classes (PO, PD and TD) have similar distributions: the stellar-mass distributions all peak at m$_{\rm star}\sim 5 \times 10^{10}$M$_{\odot}$ -- whilst the stellar mass-ratio distributions all peak at some intermediate value, dipping at low values of $\mu_{\rm star}$ (corresponding to nearly equal-mass pairs). TO pairs behave differently: both distributions decline monotonically with increasing stellar mass and mass-ratio, respectively. Pairs classified as IsoPO are biased towards very large masses. This is the reason why these pairs tend to have relatively high relative velocities (left-panel of Figure~\ref{figqphys}). They also cover a broad range of mass ratios, capable of reaching extremely large values of $\mu_{\rm star}$.  Compare to Figure~\ref{figmmrf}. 
}
 \label{figqsample}
\end{figure*}

\subsubsection{Energy Classes: Dynamics and Proximity}

The left panel of Figure~\ref{figqphys} shows the distribution of relative-velocities in our main pair sample, but now in terms of energy classes: PO (violet curve), PD (brown curve), TD (green curve) and TO (turquoise curve). For completeness, we include the pairs that do not have a third massive neighbour identified within 10 $h^{-1}$ Mpc (hereafter we use the label {\bf IsoPO} -- dotted violet curve). This figure is analogous to the right panel of Figure~\ref{figvdf}.

First, notice that the three classes involving self-bound pairs (PO, PD and TD) exhibit relative-velocity distributions that peak at similar values ($\sim$100-200 km sec$^{-1}$). The (rather mild) evolution of the location of these peaks can be explained by the degree of binding to the third neighbour. From left to right, the peaks are ordered as follows: PO (unbound to the third), PD (weakly bound to the third), and TD (strongly bound to the third). On the other hand, unbound pairs (i.e., those belonging to the TO class, which are only bound to a third galaxy) occupy an entirely different region of relative-velocity space. They tend to reach relative velocities of the order of 1,000 km sec$^{-1}$. Below, we argue that this segregation highlights the power of distinguishing galaxy pairs by energy classes. 

Recall that Section~\ref{secFlavs} suggests that the dynamics of satellite-satellite pairs are governed by the gravitational potential well of their host dark matter halo. Also, in Section~\ref{secThird} we found that a large fraction of that population consists of bound pairs (right-panel of Figure~\ref{figepair}, portion of blue curve to the left of the vertical line). Using our energy-based approach, we are capable of identifying the satellite-satellite pairs that are only bound to a third object, and not to themselves. It is reassuring to see that the relative velocity distributions corroborate this idea: TO pairs (turquoise curve) have a velocity distribution that is evidently distinct from the rest. Whilst the satellite-satellite pairs in the other energy classes are bound to each other to some degree, the TO are genuinely chance pairs -- merely tracing the gravitational potential associated to an external (third) massive galaxy in the vicinity.
 
So far, we have shown TO pairs are inherently quite different from the other energy classes: explicitly, they tend to exhibit much larger relative velocities. Let us see if we can gain additional insight with our proximity-parameter approach (Section~\ref{secThird}). The right panel of Figure~\ref{figqphys} shows the distribution of proximity parameters (equation~\ref{eqnisostar}) for our four energy classes. First, notice that TO pairs (turquoise curve) tend to inhabit regions in close proximity to a third more-massive galaxy (with the exception of a subdominant tail in the isolated regime). At the other extreme, PO pairs are more commonly found in isolation (violet curve). Surprisingly, pairs classified as PD (brown curve) also tend to live in isolation (with the exception of a subdominant tail in the near-third proximity regime). In other words, even though pairs in this class are bound to a third galaxy, the influence of the latter is rendered weak by its remoteness.

Lastly, the TD pairs exhibit a bimodal distribution of proximity parameters (green curve): with a larger peak located in the isolated regime and a smaller peak in the regime representing close proximity to the third. Notice that in proportion to the entire sample, TD pairs are evenly split between the two proximity-parameter regimes (compare green peaks to black peaks). 
Further inspection (figure not shown) reveals that the smaller peak consists of pairs associated to low-mass third neighbours ($\sim$ a few $\times 10^{9}-10^{10} M_{\odot}$), located at a sub-Mpc distances. In this case, the small distance to the third `wins' over the mass of that external galaxy, placing the pair in a region with a high value of the proximity parameter. The opposite happens to the larger peak: pairs have massive third neighbours ($M^{\rm third}_{\rm star}\gtrsim$ a few $\times 10^{10} M_{\odot}$); but these objects are located a few Mpc away from the pairs. In other words, large distance to the third `wins' over the mass of that galaxy, placing the pairs in isolation relative to their corresponding third neighbouring galaxy.

\subsubsection{Energy Classes: The Sample}

For completeness, Figure~\ref{figqsample} shows the distribution of stellar masses (left panel) and stellar mass-ratios (right panel) for each of our four energy classes.  This is analogous to Figure~\ref{figmmrf}, discussed in Section~\ref{secflavsample}. Notice that the first three classes (PO, PD and TD -- all involving self-bound pairs) behave similarly. In terms of mass, they all exhibit a pronounced peak at $\sim 5\times 10^{10} M_{\odot}$. Likewise, the distributions of stellar mass ratios behave similary for these three classes: they all peak at some intermediate value and plummet at high and low values of $\mu_{\rm star}$. 

On the other hand, the pairs belonging to the TO class display rather distinct trends. In this case, both distributions (turquoise curves in both panels) decrease monotonically with increasing $m_{\rm star}$ and $\mu_{\rm star}$, respectively. Surprising, recall that our analysis of the relative velocity and proximity parameter (Figure~\ref{figqphys}) also reveal a segregation of the pairs belonging to the TO class from the rest. This is a rather remarkable result. Namely, we have demonstrated that our methods are capable of {\it identifying those galaxy pairs ($\sim$30\%) that are merely chance associations}. Further, we speculate that these pairs actually have absolutely no chance of ever merging.

Lastly, the subset of bound pairs with no third neighbouring galaxy within 10 $h^{-1}$ Mpc (IsoPO -- violet dotted curves) is biased towards more massive systems: peaking at $\sim 2\times 10^{11} M_{\odot}$. Recall that, in general, galaxy pairs where one of the members is a very massive galaxy have a broader range of mass ratios accessible to them (right panel of Figure~\ref{figsample}). This is true also for the IsoPO pairs: the distribution of mass-ratios for this particular classification is rather flat. Moreover, it is natural to expect massive pairs in extreme isolation to display large velocity differences. This is corroborated by the left panel of Figure~\ref{figqphys}. However, in stark contrast to TO pairs, whose high velocities are driven by an external agent -- the high velocities experienced by IsoPO pairs here are actually driven by the massive galaxies themselves, which happen to be in extreme isolation (i.e., they do not have a more massive neighbour within 10 $h^{-1}$ Mpc). 

\begin{figure*}
 \centering
  \includegraphics[width=\columnwidth]{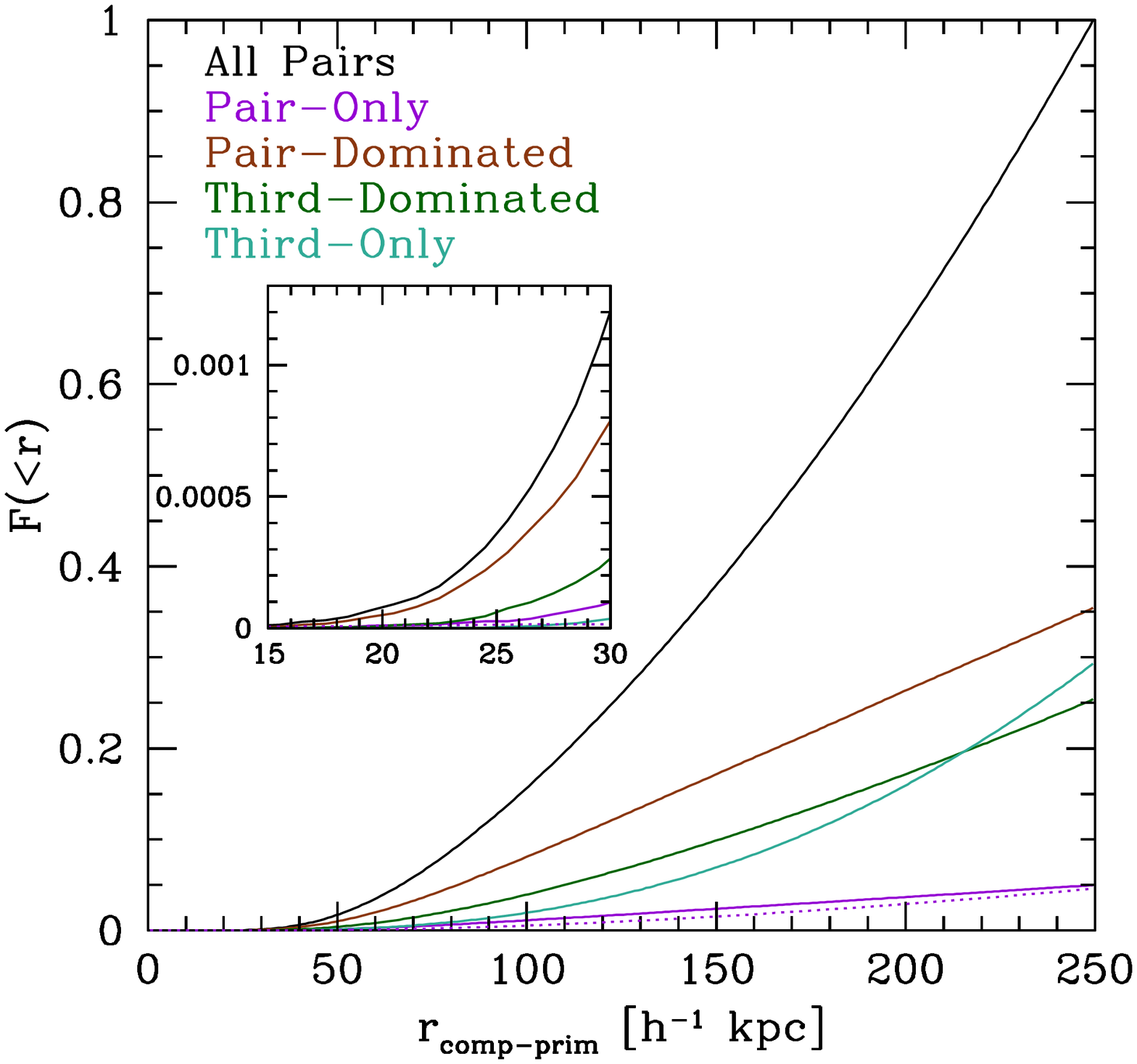}
  \includegraphics[width=\columnwidth]{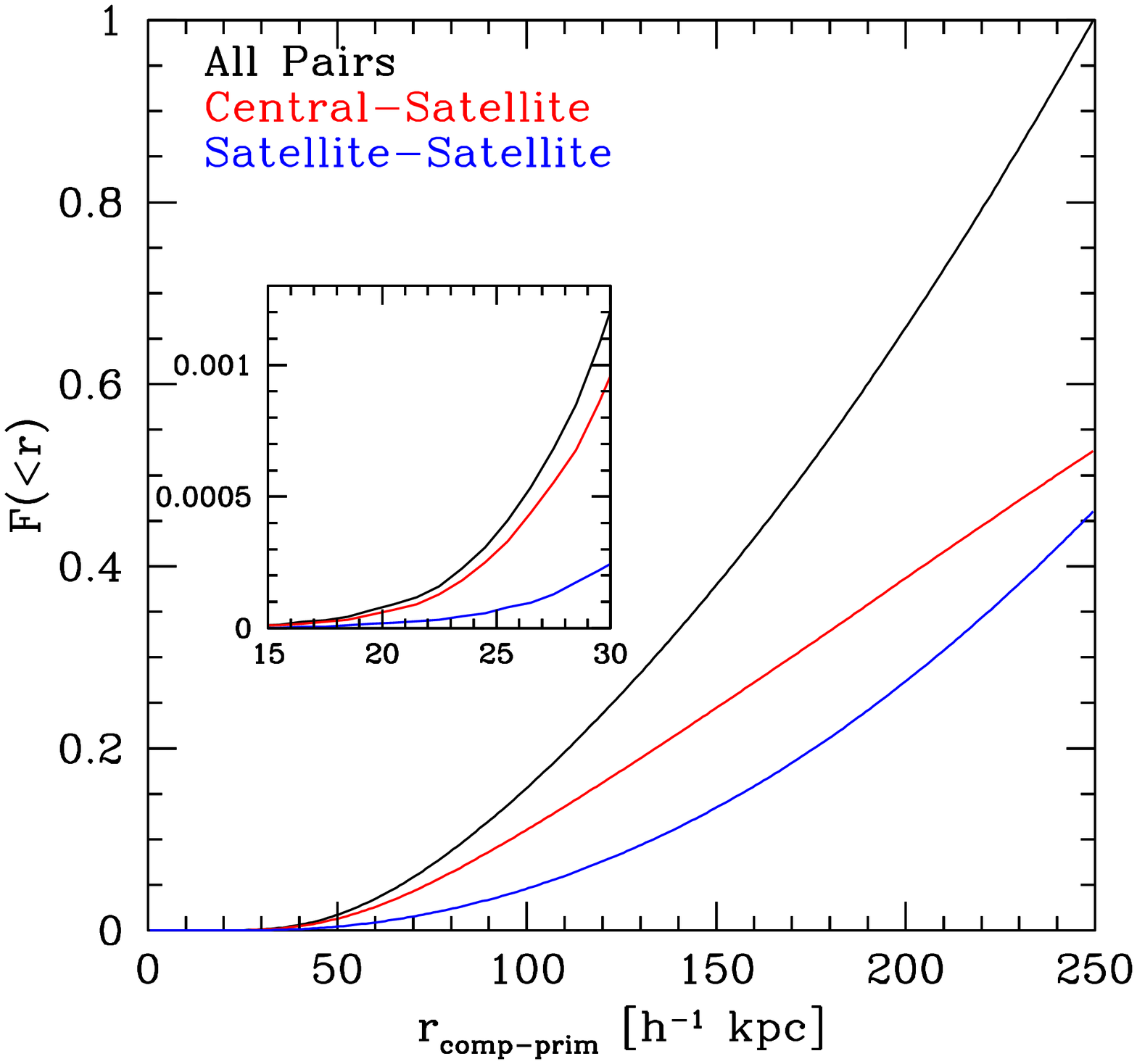}
 \caption{Cumulative distribution of separations (normalised to unity at 250 $h^{-1}$ kpc -- {\it lower panels}); in terms of energy classes ({\it left}); and in terms of galaxy pair flavours ({\it right}). 
The insets show the cumulative distributions in the extremely low separation regime: $r<30$ $h^{-1}$ kpc. The black curves represent the full sample. In the cumulative distributions, pairs classified as PO (violet), PD (brown), TD (green) or central-satellite (red); whilst pairs classified as TO (turquoise) or satellite-satellite (blue) increase faster than linear. The insets demonstrate that at the smallest separations (below 30 $h^{-1}$ kpc), the samples are dominated overwhelmingly by PD (and central-satellite) pairs, followed by TD pairs. See Table~\ref{tablecumd} for actual numbers using various separation-constrained subsets.
}
 \label{figcumd}
\end{figure*}

\section{Discussion}\label{secDiscussion}

\subsection{An Emerging Picture}

So far, the principal aim of this paper has been to investigate the dynamics of galaxy pairs, and their connection to their environment. Three platforms have been explored: {\it flavour}, {\it proximity} (relative to a third massive neighbour), and {\it binding parameters}. We devote this section to briefly summarize the coherent picture that emerges by employing these three complementary approaches.

First, our notion of galaxy-pair flavour (i.e., dark matter halo membership) is powerful because it connects well to models of galaxy formation, which treat central and satellite galaxies in different ways (i.e., satellites are more susceptible to environmental effects, such as ram-pressure stripping, etc.). In particular, we find that {\it central-satellite} pairs exhibit a tight correlation between relative velocity and mass; a correlation that does not exist in the case of {\it satellite-satellite} pairs (see Figure~\ref{figvmf}).

Second, our proximity approach connects well with observations. Namely, projection effects aside, it is relatively straightforward to spot the closest/most-massive object in the surroundings of a galaxy pair identified in a survey (Figure~\ref{figsloan}). Moreover, our proximy criteria (equations~\ref{eqnisodyn} and \ref{eqnisostar}) can separate our sample into two clearly discernible sub-populations: {\it isolated} pairs (with $\Phi^{\rm third}_{\rm star}<\Phi_{11}$), and pairs in {\it close proximity} to a third massive neighbour (with $\Phi^{\rm third}_{\rm star}>\Phi_{11}$ -- as illustrated by Figure~\ref{figsloan}). In the context of flavours: the majority of central-satellite pairs tend to be isolated, whist most satellite-satellite prefer to be in close proximity to their corresponding third galaxies (see Figure~\ref{figiso}).

\begin{center}
\begin{table*}
\begin{tabular}{|| l |l l || l || l || l || l ||}
\hline
Class/Flavour & $\leq$30 $h^{-1}$ kpc & $\leq$50 $h^{-1}$ kpc & $\leq$100 $h^{-1}$ kpc & $\leq$250 $h^{-1}$ kpc & Colour \\
\hline \hline 
Pair-Only (PO+IsoPO) & 9.3\% & 10.6\% & 10.6\% & 9.6\% & Violet \\
 & $=$ 7.9\% $+$ 1.4\% & $=$ 8.1\% $+$ 2.5\% & $=$ 7.1\% $+$3.4\% & $=$ 5.0\% $+$ 4.6\% &  \\
Pair-Dominated (PD) & 66.7 \% & 59.0\% & 51.7\% & 35.4\% & Brown\\
Third-Dominated (TD) & 21.1\% & 23.1\% & 25.5\% & 25.4\% & Green\\
Third-Only (TO) & 2.8\% & 7.3\% & 12.4\% & 29.3\% & Turquoise\\
\hline 
Central-Satellite & 79.6\% & 75.7\% & 70.7\% & 52.7\% & Red\\
Satellite-Satellite & 20.4\% & 24.3\% & 29.3\% & 46.1\% & Blue\\

\hline
\end{tabular}
\caption{Separation-constrained subsets: relative contributions from each energy class and galaxy-pair flavours. Each row refers to a specific energy class (top four rows) or flavour (bottom two rows). Each column represents a subset of the main galaxy pair sample, constrained to have separations below the indicated value. The numbers represent the percentage of pairs in a given (separation-constrained) subset that belong to a given energy-class/flavour. The last column indicates the colour used in the figures. 
}
\label{tablecumd}
\end{table*}
\end{center}

Third, our binding parameters (equations~\ref{eqnbpair} and \ref{eqnbthird}) hold clues to the physics driving the dynamical evolution of the pair. Surprisingly, this approach unveils a more complex underlying picture. First of all, a vast majority of central-satellite pairs are self-bound. Likewise, satellite-satellite pairs tend to be bound to a third massive neighbour. However, a large fraction of satellite-satellite pairs tend to be self-bound, whilst the great majority of central-satellite pairs are {\it also} bound to a third galaxy. In other words, most pairs in our general sample are both self-bound, and bound to a third object.
This simple result highlights the fact that the connection between pairs of galaxies to their surroundings is far from trivial -- thus, stressing the need for investigations like the one presented in this paper.

This energy-based formalism allows us to classify our galaxy pairs into four energy classes: Pair-Only (PO), Pair-Dominated (PD), Third-Dominated (TD), and Third-Only (TO). It is tentative to speculate that, with energetics, one can potentially predict the future merger fate of these pairs. Namely, it is plausible that the galaxies in pairs belonging to the PO class, which are bound and isolated, are very likely to merge with each other. Likewise, pairs in the TO class, which are unbound chance pairs bound to the halo hosting a third massive galaxy, are very unlikely to merge. 

The situation might be slightly more complicated for those pairs which happen to be both self-bound, and bound to a third neighbour. In that case, we speculate the following. PD pairs are likely to merge as well, but the type of orbits they experience will be modified by the presence of a third neighbour \citep[as seen in numerical simulations by][]{martig08}. Galaxies in TD pairs, on the other hand, are likely to interact with one another, but will eventually merge with their third neighbour \citep[as demonstrated by][for many pairs]{moreno12}.

At this point, we stress that at this stage, these claims connecting energetics to merging are merely conjectures. Actual verification involving the use of merger history trees to directly track of the orbits, is the subject of an upcoming paper (Moreno et al., in prep).

\subsection{Extremely Close Pairs}\label{seccumd}

So far, we have analysed pairs of galaxies selected to have separations under 250 $h^{-1}$ kpc (in co-moving three-dimensional space). However, for certain applications, it might be more suitable to constrain our sample to a more limited range of separations. For instance, over the years, some works that use close pairs to estimate the galaxy merger rate require the two galaxies to have projected separations under $\sim$30 kpc \citep[e.g.,][]{patton97,patton00,patton08,rawat08,ellison08,bluck09,man12,bluck12}. However, other works have found that interaction-induced effects, such as star-formation enhancement, appear out to projected separations of $\sim$100 kpc \citep{ellison08}.  Moreover, recent evidence has shown that the orbital extent of these effects can reach as far out as $\sim$150 kpc in projected separation \citep{patton13}. 

Inspired by these observational results, we devote this section to a brief discussion on galaxy-pair subsamples with more strict (three-dimensional) proximity criteria. For the sake of brevity, we only explore how the contributions of different energy classes and flavours change as we modify the maximum separation pairs can have. Other (possibly more interesting) applications are deferred to future works.

Figure~\ref{figcumd} shows the cumulative distributions (normalised to unity at 250 $h^{-1}$ kpc) for the different energy classes (left panel) and flavours (right panel). 
First notice that, for the energy classes involving bound pairs (PO, PD, TD and IsoPO), the cumulative distributions increase in an almost linear fashion (or possibly faster for TD pairs). The same is true for central-satellite pairs. The cumulative distributions for TO pairs (and for satellite-satellite pairs) appear to increase in a faster-than-linear fashion. For instance, at separations of $\sim 170$ $h^{-1}$ kpc, the TO pairs begin to `catch up' with (and increase faster than) the PD pairs. Lastly, note that PD and central-satellite pairs dominate the budget across every separation regime.

The annexed insets zoom into the regime containing the most {\bf extremely close pairs}: at $r\leq30$ $h^{-1}$ kpc, only six times larger than the softening scale of the Millennium Simulation (see Section~\ref{subsecCaveats} for a discussion of potential issues that may arise as we approach this separation regime). In this regime, all the cumulative distributions increase in a non-linear fashion. Focusing on small scales, notice that the contribution from PD pairs (brown) increases from $\sim$35\% at 250 $h^{-1}$ kpc up to $\sim$ 67\% at 30 $h^{-1}$ kpc. Similarly, central-satellite pairs (red curves) increase from $\sim$53\% up to $\sim$80\%. In contrast, the contribution from pairs classified as TO (turquoise) decrease from $\sim$30\% down to $\sim$3\%; whilst the satellite-satellite contributions decreases from $\sim$46\% down to $\sim$20\%. On the other hand, TD pairs (green) decrease in a smoother fashion ($\sim$25\% $\rightarrow \sim$21\%). Lastly, PO pairs (violet) increase ($\sim$5\% $\rightarrow \sim$8\%) whilst IsoPO pairs (dotted-violet) decrease ($\sim$4.6\% $\rightarrow \sim$1.4\%). See Table~\ref{tablecumd} for actual values in relation to the following restricted subsets: $r\leq\,$30, 50, 100 and 250 $h^{-1}$ kpc.

It is reassuring to see that {\it TO pairs are largely non-existent at $r\leq$\,30 $h^{-1}$ kpc} ($\sim$2.8\%), even though they are the second most dominant class at $r\leq$ 250 $h^{-1}$ kpc. In other words, as far as merger studies that utilize close pairs are concerned, it is important to know that the contribution from chance associations is negligible at small separations.  However, once the pairs classified as TD are included to the budget of pairs energetically-dominated by an external third massive galaxy (i.e., TD plus TO), this fraction rises to $\sim$24\% of the total at 30 $h^{-1}$ kpc. At this point we can speculate that, in stark contrast to what is often assumed in the literature, our results suggest that about a quarter of all the extremely close galaxy pairs (with $r\leq$\,30 $h^{-1}$ kpc) may, in fact, never merge -- either because they are merely chance associations orbiting a massive neighbouring galaxy or because they are weakly bound. Of course, this remains to be demonstrated directly with actual merger history trees (Moreno et al., in prep).

\subsection{Caveats and Future Directions}\label{subsecCaveats}

In this section, we discuss the assumptions we adopt in this work, and possible ways to improve our methods. We also highlight possible avenues for future investigations.

\subsubsection{Our Cosmological Simulation}
First, we recognize that, despite its successes over the years, the Millennium Simulation is limited in spatial and mass resolution. A simulation with higher resolution would allow us to (i) probe separations below 15 $h^{-1}$ kpc (insets in Figure~\ref{figcumd}), and (ii) track subhaloes for longer times, down to stages immediately before merging. However, we do not expect these particular changes to alter the trends reported here in any dramatic way.

On the other hand, higher resolution would also allow us to probe lower masses and a larger diversity of mass ratios (Figure~\ref{figsample}). For instance, access to lower masses would permit investigations involving pairs of dwarf galaxies \citep{besla12,fattahi13,gonzalez13}. However, the applicability of our abundance-matching technique is not clear in that regime -- and thus a study of dwarf galaxy pairs would require us to employ a physical model instead \citep[e.g.,][]{starkenburg13}. Similarly, higher mass ratios would in principle motivate better studies related to minor mergers \citep{woods06,lopezsanjuan10,bluck12,edwards12,newman12}. Nevertheless, a simulation with higher resolution would most likely sacrifice volume, limiting the number of massive objects severely. For our present purposes, the scope available to the Millennium Simulation suffices.

Often authors who employ the Millennium Simulation keep track of `orphan' galaxies \citep{guo11,moster13}, which are defined as those galaxies whose host subhaloes have gone below the resolution of the simulation. The effect of ignoring this population is that we might be missing some satellite galaxies. However, our results depend strongly on knowing the location of these satellites, and their dynamical masses. Moreover, our 100-particle condition at infall automatically discards the majority of subhaloes that are more susceptible to tidal-stripping by their host \citep[as shown by][]{moreno12}. For these reasons, we ignore orphan galaxies altogether in this work.

The ability to track subhaloes as they orbit the densest cores of massive dark matter haloes also has an impact on the number of substructures identified in the simulation, and their subsequent merger history. For instance, \cite{behroozi13} suggest that, in relation to their phase-space temporal halo finder, \textsc{subfind} (the halo finder adopted by the Millennium Simulation) may under-predict the number of major mergers within 30 $h^{-1}$ kpc. However, given that this degree of separation is already quite close to the resolution of the simulation, in this work we make no strong claims below this threshold (see Figure~\ref{figcumd} and associated discussion). Furthermore, the halo-finder comparison by \cite{onions12} demonstrates that most subhalo-finding algorithms in the literature yield similar values for the cumulative mass in subhaloes as a function of distance from the centre of the host -- with disagreements appearing only once the regime below 30 $h^{-1}$ kpc is reached; a regime that we choose to ignore here altogether.

\subsubsection{Our Simulated Galaxies}

Another possible point of contention is the use of abundance-matching techniques to link galaxies to subhaloes \citep{moster13}.
First, we should mention that there are other abundance-matching techniques available in the literature, each with its own advantages and limitations \citep{conroy09,moster10,behroozi10,trujillo11,yang12,rodriguez12,rodriguez13,reddick13}. However, the majority of these works are designed for precision-cosmology purposes. For our current investigations, the small variations present in those models are unlikely to change our conclusions in any significant way. Therefore, it is reasonable to commit to a single abundance-matching framework; and the model of \cite{moster13} not only meets our needs, but is already based on the Millennium Simulation.

One could also argue that we should use semi-analytic models (SAMs) of galaxy formation instead, some of which are also already based on the Millennium Simulation \citep{bower06,delucia07,guo11}. Similarly, we could employ existing cosmological simulations with hydrodynamics \citep{dimatteo08,crain09,schaye10,tonnesen12,dimatteo12,muldrew13}. However, our work focuses exclusively on the dynamics of galaxy pairs. Thus, for our present purposes, the details characterizing galaxies in our sample (beyond stellar mass) are largely irrelevant.

In the long term, our ultimate goal is to investigate the effects of interactions on galaxies (i.e., star-formation enhancement, etc.) in a cosmological context. By construction, SAMs do not include these effects at the earliest stages of interaction (although recipes describing post-merger galaxies are quite developed). Likewise, these phenomena have failed to appear in some large-volume cosmological hydro-simulations \citep{tonnesen12} -- even though simulations with smaller volumes and higher resolution do exhibit them \citep{tissera02,perez06,tissera12}. It remains to be seen if the next generation of large-volume/high-resolution cosmological simulations with hydrodynamics (i.e., Illustris, Eagle, etc.) are capable of properly capturing the physical processes relevant during the earliest stages of merging (M. Vogelsberger \& C.~S.~Frenk, private communications).  Either way, direct comparison with the observational results produced (i.e., by our group and others) in recent years would be very fruitful. 

\subsubsection{Our Galaxy Pairs and their Third Neighbour}

Our selection criterion, based on a fixed separation threshold, is the simplest way  to construct a galaxy pair catalogue. As pointed out by \cite{tonnesen12}, one shortcoming of this choice is that one could in principle miss bound pairs at separations larger than our adopted threshold. However, it is not clear that this situation will be ameliorated in observational studies, so we opt not to pursue this any further.

Another option is to require pairs to belong to the same dark matter halo. By construction, this would only select flavours (a) and (b) in Table~\ref{tableflavs}. Although we did, in fact, find that these two flavours dominate the sample; this is only true at $z=0$. At higher redshift, the progenitors of our pairs are likely to have smaller virial radii -- and therefore be central-central pairs. In an upcoming paper (Moreno et al., in prep), we plan to quantify the transition from one flavour to another. At this stage, however, we prefer to keep our current selection scheme. 

Also, one could aim to select those pairs that are more susceptible to tidal effects. Some authors have tried to quantify this via a so-called `tidal index' \citep[e.g.,][]{verley07,sabater13}. However, the physical connection between close encounters and tidal effects on galaxies is still not well understood. It would surely be interesting to identify these effects in cosmological hydro-simulations, but this is certainly beyond the scope of this work. Another limitation with the tidal-index approach is that the visibility time for a high-index encounter might be too short.  In other words, just because a pair has a high tidal index at pericentre does not mean that it will continue to have a high value at larger separations (near apocentre). This limitation is not unique to this method, but is also present in proximity-based methods like ours. Namely, a widely separated pair might have had a close encounter in the past \citep[as demonstrated by][]{moreno12}; but would not pass our selection criteria. 

We note that, in our criteria, a given galaxy can be a member of more than one pair (e.g., a galaxy in a triplet can potentially participate in two pairs) -- whilst other authors prefer to select only the nearest galaxy as the companion. We adopt the former approach for technical convenience only, as pairs selected with alternative, `single-companion' methods, can always be extracted from our main sample.

Lastly, we briefly comment on our definition of the third neighbour. Recall that in this work, the third galaxy is selected such that it has the greatest value of ${M^{\rm third}_{\rm dyn}}/{R_{\rm pair-third}}$ --  where ${M^{\rm third}_{\rm dyn}}$ is the dynamical (dark matter) mass of any neighbour {\it more massive} than either member of the pair (in terms of dynamical mass), and ${R_{\rm pair-third}}$ is the distance from the centre-of-mass of the pair to that object. This definition is adopted because, in this paper, we are interested in investigating what is driving the dynamics of galaxy pairs. 

However, we recognize that the existence of `third' galaxies with masses {\it below} those of either member of the pair, is also important. For instance, a third nearby galaxy might signal that the pair is actually embedded in a triplet \citep[e.g.,][]{darg11,omill12,duplancic13} or a compact group \citep[i.e.,][]{mcconnachie08,brasseur09,mcconnachie09,diaz10,mendel11}. 

Moreover, the hydrodynamical simulations of \cite{moster12} also show that, in general, binary mergers do not evolve alone. Namely, the merger between two galaxies is commonly interrupted by a third low-mass intruder. These results are in line with the main message in this paper: {\it it is not always realistic to treat merging galaxies as if they were complete isolation from the rest of the Universe.} 

Finally, our results suggest that looking at only the third galaxy suffices to identify the main drivers governing the dynamics of galaxy-galaxy pairs. For this reason, we do not explore any additional neighbours (i.e., a fourth, a fifth, etc., galaxy in the vicinity of a pair) in this paper. Moreover, we use the notion of third massive neighbour to quantify the environment of a given pair. In particular, our results show that the distribution of proximity parameters is strongly bimodal (right-hand panel of Figure~\ref{figiso}). It would be interesting to export this analysis to {\it single} galaxies embedded in a proper cosmological context. For instance, one could investigate the location of a galaxy's nearest/more-massive neighbour (without imposing a strict separation cut), and verify if these two objects are bound to each other. Such generalizations will be the subject of a future study.

\subsubsection{Connection to Observations}

In this work, we select galaxy pairs by using a proximity condition in three dimensions. However, measuring the three-dimensional separation between two galaxies in the sky is almost impossible. Instead, observers rely on projected separations and line-of-sight velocity differences to select close pairs. In an upcoming paper (Moreno et al., in prep), we plan to construct a close pair catalogue based on a mock galaxy survey, by following the same steps an observer would take \citep[e.g., Figure~1 of][]{hayward13}. Such a study is analogous to the work of \cite{kitzbichler08}, but it will evaluate the influence of environment \citep[in this context, see also][]{jian12}. Moreover, this work will not be limited exclusively to merging galaxies, but will incorporate all types of galaxy pairs (i.e., fly-bys).

Additionally, we plan to investigate other ways to quantify the environment surrounding pairs of galaxies. For instance, we are curious to see if the notion of third neighbour remains clear when projection effects are introduced. Likewise, we wish to explore the language of flavours in a mock survey, where dark matter haloes are identified with group catalogues \citep{yang07}. Other metrics -- such as the distance to the fifth nearest neighbour or Voronoi tessellations \citep{scoville13} -- can also be adopted.

Finally, one could argue that a more observationally-motivated mock pair sample should have been utilized from the beginning of this paper. The problem with that approach, however, is that the introduction of projection effects and interlopers could mask the gravitational processes driving the dynamics of galaxy pairs. For this reason, it is more sensible to explore these configurations from a three-dimensional point of view first, before introducing all the complications a virtual observer could face.

\subsubsection{Galaxy Pairs as Tracers of Future Mergers}

In this paper, we use energetics and proximity to a more massive neighbour to quantify how galaxy pairs are embedded in their respective environments. However, it would be interesting to verify whether or not these metrics can tell us something about the future merging fate of these pairs of galaxies. For instance, it is sensible to speculate that galaxies in pairs classified as PO will {\it definitely} merge with each others, whilst those in TO-pairs will {\it definitely not}. Similarly, one could say that PD-pairs are {\it very likely} to merge, whilst those classified as TD are {\it very unlikely} to do so. In the last two cases, we anticipate the presence of a third massive neighbour to play a very important role. 

To address this, in an upcoming paper we will analyze galaxy pairs selected at a high redshift, and track their merger histories forward in time (Moreno et al., in prep). In particular, we will evaluate whether or not being a member of a given energy class tells us something about the probability that two galaxies in close proximity will merge with each other. We will also quantify the amount of time a galaxy pair `spends' in a given energy class. In other words, we can answer questions of the following sort: {\it What fraction of pairs can be treated as being in complete isolation for 1 Gyr (or any other longer timescale)?}

There are other ways in which detailed hydrodynamical studies of two merging galaxies can benefit from cosmological simulations. For instance, \cite{khochfar06} use the latter type of simulations to extract the appropriate orbital parameters (i.e., pericentric distances, circularities) needed to model major galaxy mergers. However, by construction, that work only takes into account pairs that will definitely merge in the future. In contrast, our work incorporates all galaxy pairs (within a given separation threshold), more in line with what observers measure in surveys. Similarly, our framework can quantify the influence of a massive neighbour on the specific orbital parameters of a given galaxy pair -- further refining the estimates provided by \cite{khochfar06}.

To our knowledge, detailed investigations on the evolution of galaxy pairs in a truly cosmological setting are still lacking. The need for such studies is very important, as demonstrated by controlled numerical simulations of merging galaxies embedded in clusters. For example, \cite{mihos03} simulate an equal-mass merger immersed in a Coma-like gravitational potential. Their main result is that the tidal field of the cluster diminishes the longevity (and detection ability) of tidal debris. However, they also find that, by imparting energy to the orbits of the two galaxies, the tidal field prolongs the time it takes them to merge with each other. Beyond this particular configuration, it is possible for the merger to be delayed indefinitely -- in other words, to have a pair of galaxies that are in close proximity, but never merge \citep[in line with the findings of][]{moreno12}.

On a similar vein, \cite{martig08} analyze a suite of merger simulations embedded in group and cluster environments. These authors find that the external potential is capable of producing gravitational focussing, which may accelerate the merging process. Moreover, they show that interaction-induced star formation enhancement is systematically stronger in clusters than in the field. Interestingly enough, the actual intensity and timing of this enhancement is strongly dependent on the chosen orbital configuration of the merging galaxies within the cluster. For this reason, a comparison between this work and that of  \cite{mihos03} is not straightforward -- further highlighting the need to execute such studies in a cosmologically-motivated setting.

The suggestion that, for some orbital configurations, interaction-induced star formation enhancement can be stronger in the vicinity of clusters than in the field, is very promising. It is standard for semi-analytic treatments of galaxy formation to stop or delay star formation as soon as a galaxy becomes a satellite \citep[e.g.,][]{taranu12,wetzel12,wetzel13}. However, the simulations of \cite{martig08} show that this situation is far more complex. Namely, close encounters are more effective in clusters, and thus more conductive to starburts. Indeed, tentative evidence of this has been found observationally in the outskirts of galaxy clusters \citep{heiderman09,mahajan12}. Theoretical studies in this direction can be addressed with upcoming cosmological hydro-simulations (i.e., Illustris/Eagle), and will be the subject of future investigations.

Lastly, we also envisage this kind of cosmologically-motivated studies to be relevant to systems in the local Universe.  For instance, it could help us describe systems like the Magellanic Clouds in relation to the Galaxy \citep{besla12}; or close galaxy pairs embedded in the Virgo Cluster (the subject of an upcoming project; L. Ferrarese -- private communication). 

\section{Summary}\label{secConclusions}

Undoubtedly, galaxy mergers embody a fundamental pillar in modern extragalactic astronomy. A testament to this is the large number of numerical efforts aimed to describe the merging process in detail. Unfortunately, idealised galaxy-merger simulations assume that merging galaxies evolve in isolation. This paper demonstrates that this commonly used assumption is not entirely correct. Namely,  the rest Universe has inevitable effects on the dynamics of galaxy pairs. See Figure~\ref{figsloan} for an illustrative example involving two discs in the presence of a massive elliptical in their vicinity.

Therefore, to investigate this problem, we use the Millennium Simulation \citep{springel05}, along with a multi-epoch abundance matching technique \citep{moster13}. We construct a catalogue containing 1,336,289 simulated pairs of galaxies, selected to have three dimensional separations lower than 250 $h^{-1}$ kpc,  and mass at infall higher than $8.6 \times 10^{10} h^{-1}$ M$_{\odot}$ (corresponding to 100 particles).

In Section~\ref{secFlavs}, we introduce the notion of {\it flavour}, which classifies pairs in terms of membership to their corresponding dark matter haloes (Figure~\ref{figflavs}). We find the following:

\subsection{Galaxy Pair Flavours}

\begin{itemize}

\item The majority of our pairs are either {\it central-satellite} pairs ($\sim$52.7\%) or {\it satellite-satellite} pairs ($\sim$46.1\%) in a single dark matter halo. The contribution from other combinations, involving two haloes, is negligible (Table~\ref{tableflavs}).

\item Central-satellite pairs exhibit a tight correlation between relative velocity and total stellar mass. This suggest that, in these cases, the dynamics are primarily driven by the galaxies in the pairs. 

\item For satellite-satellite pairs, there is no correlation between relative velocity and stellar mass. Instead, their dynamics is driven by their host dark matter halo.

\item The distribution of relative velocities peaks at $\sim$200 km sec$^{-1}$ for central-satellite pairs, and at $\sim$600 km sec$^{-1}$ for satellite-satellite pairs.

\end{itemize}

\subsection{Third Neighbour}

In Section~\ref{secThird}, we characterize the environment around a galaxy pair by identifying the nearest/most-massive {`\it third'} object in the vicinity (equations~\ref{eqnisodyn} and \ref{eqnisostar}). We find that:

\begin{itemize}

\item 57\% pairs live in isolation, whilst the remaining 43\% inhabit near-third proximity regions (Figure~\ref{figiso}).

\item Central-satellite pairs tend to be in isolation (94\%), whilst satellite-satellite pairs prefer to be in close proximity to a third neighbour (91\%).

\end{itemize}

\subsection{Binding Energy}

In Section~\ref{secEnergy}, we analyse pairs from an energy-based perspective (equations~\ref{eqnbpair}, \ref{eqnbthird} and associated text). We find the following results:

\begin{itemize}

\item In our sample, 70\% of the pairs are bound (Figure~\ref{figepair}).

\item Essentially all central-satellite pairs are bound ($\gtrsim$98\%). However, only 62\% of satellite-satellite pairs are unbound; the rest (38\%) correspond to bound configurations. 

\item Satellite-satellite pairs tend to be more strongly bound to their third neighbours than their central-satellite counterparts (Figure~\ref{figethird}).

\item Essentially, all satellite-satellite pairs are bound to a third neighbour ($\sim$99\%). Surprisingly, most central-satellite pairs are also bound to their third neighbour too (83\%).

\item In fact, most pairs in the general sample are bound to a third massive/nearby galaxy (91\%).

\item Only 9\% of the pairs are truly in energetic isolation. These systems are precisely the ones that most closely resemble galaxies in idealised merger simulations.

\end{itemize}

\subsection{Energy Classes}

In Section~\ref{secPairThird}, we compare the strength of binding within a pair versus the binding relative to their third neighbour (Figure~\ref{figbb}). We find that:

\begin{itemize}

\item A large majority of paired galaxies are bound to both their companions and a third neighbouring galaxy (61\%).

\item Our binding-energy approach allows us to separate galaxy pairs into four distinct {\it energy classes}: Pair-Only (PO -- 9\%), Pair-Dominated (PD -- 35.4\%), Third-Dominated (TD -- 25.4\%) and Third-Only (TO -- 29.3\%).

\item The dynamical properties of pairs classified as TO are quite distinct from the other three energy classes. Namely, pairs belonging to this class (i) exhibit higher relative velocities; (ii) are primarily in close proximity to a third galaxy; and (iii) are biased towards low stellar masses and (`major') mass ratios (Figures~\ref{figqphys} and \ref{figqsample}).

\end{itemize}

\subsection{Extremely Close Galaxy Pairs}

Lastly, in Section~\ref{seccumd}, we explore the consequences of placing more strict proximity conditions (Figure~\ref{figcumd}). Here are our main findings:

\begin{itemize}

\item In the extremely close regime (with $r\,\leq\,$30 $h^{-1}$ kpc), the relative contributions of our four energy classes are: PO (9.3\%), PD (66.7\%), TD (21.1\%) and TO (2.8\%).

\item At low separations, the contributions of PO, PD and central-satellite pairs increase, whilst the contributions of TD, TO and satellite-satellite pairs decline (Table~\ref{tablecumd}).

\item At low separations, TO are negligible (2.8\%), but the combined contributions of TO and TD account for $\sim$24\%. In other words, almost a {\it quarter} of all pairs at small separations are either unbound or weakly bound. We speculate that these galaxy pairs have little chance of merging!

\end{itemize}

We make the conjecture that energetics can in principle predict the future merging history of a pair. An upcoming paper will utilize merger history trees to test this idea. Also, upcoming work will employ mock surveys to quantify the impact of using projected separations, line-of-sight velocity differences, and other complications observers commonly encounter with real galaxy surveys (Moreno et al., in prep).

\section*{acknowledgments}

The authors thank Andrew Benson, Chris Hayward, Florent Renaud, Ravi Sheth, and Else Starkenburg for comments on an earlier draft -- and the reviewer, Fr\'{e}d\'{e}ric Bournaud, for a very useful and timely report. JM acknowledges partial funding from the Canadian Institute for Theoretical Astrophysics. JM, AB, SE and DP are funded by the Natural Sciences and Engineering Research Council of Canada.


\bibliographystyle{mn2e}

\label{lastpage}

\end{document}